\documentclass[aps,pre,preprint,superscriptaddress]{revtex4-2}

\usepackage{graphicx}
\usepackage{dcolumn}
\usepackage{bm}
\usepackage{amsmath,amssymb}
\usepackage{xcolor}
\usepackage[colorlinks=true,allcolors=blue]{hyperref}

\usepackage{subcaption}
\usepackage{mathrsfs}

\begin{document}
\title{Self-phoretic oscillatory motion in a one-dimensional channel}
\author{Leah Anderson}
\affiliation{Université Bordeaux, CNRS, LOMA, UMR 5798, F-33400 Talence, France}
\author{David S. Dean}
\affiliation{Université Bordeaux, CNRS, LOMA, UMR 5798, F-33400 Talence, France}

\date{\today}
\begin{abstract}
We study a simple model for a particle that is active due to self-phoresis and that has been proposed to model symmetric camphor grains. The particle generates a concentration field through the continuous emission of a chemical substance, and its motion is driven by gradients of this field as it diffuses within a confined channel whose ends perfectly reflect the chemical. The reflection of the chemical field leads to an effective confinement of the particle, which itself is reflected before encountering the channel ends. The system displays a transition from a passive state, where the particle rests at the channel midpoint, to an active state characterized by highly regular, non-chaotic oscillations. We analytically construct the phase diagram and derive the oscillation frequency and amplitude in the vicinity of the transition. A perturbative analysis perfectly describes the dynamics of the particle even for oscillations as large as half the channel size. Furthermore, we develop an analysis which explains the mechanism of particle reflection close to the channel edges in the regime of large activity.
\end{abstract}
\maketitle

\section{Introduction}
An important class of active matter \cite{reviewSPP} are systems where particle self-propulsion is driven by phoretic forces arising from gradients in a surrounding field - a field that is generated by the particles themselves. This self-interaction phenomenon also appears in other areas of physics, most notably in relativistic electrodynamics through the Abraham-Lorentz-Dirac force \cite{Abraham, Dirac}, the first historical example of a self-force. This force describes the effect on an accelerating charged particle due to its own emitted electromagnetic radiation. A particularly important case of self-propelled motion in biology is chemotaxis, in which microorganisms navigate along self-generated chemical gradients to avoid toxins or seek nutrient-rich regions \cite{microorg1, microorg2}. In addition to biological self-propelled particles, various synthetic active “swimmers” have been engineered to harness forces arising from self-generated field gradients \cite{Syntheticswimmers}. Notable examples include asymmetric camphor boats \cite{Camphorboat1, Camphorboat2} and Janus particles \cite{JanusParticle1, JanusParticle2}, both of which emit non-uniform chemical fields as a direct consequence of their structural asymmetry. If the propulsion mechanism of such systems arises from chemical concentration gradients, it is termed self-diffusiophoresis. The self-generated field gradient may also be thermal in origin, in which case the mechanism is termed self-thermophoresis \cite{Anderson}.

We adopt the simplest one-dimensional model for an overdamped point particle linearly coupled to its self-emitted chemical field, $c(x,t)$.   The spatial gradient of the chemical field, denoted by $c'(x,t)$, then induces a linear response velocity for the particle given by
\begin{equation} 
\label{0}
    V_t = \frac{dX_t}{dt}=- \lambda \ c' ({X}_t,t).
\end{equation}
The coefficient $\lambda$ characterizes the strength of the particle–field coupling and is often referred to as the phoretic mobility. A negative $\lambda$ corresponds to chemoattraction in a biochemical context, where the particle is drawn toward regions it has previously visited. In one and two dimensions, this behavior has been shown to lead to self-trapping \cite{TsorideGennes}. In contrast, a positive $\lambda$ causes the particle to avoid previously visited regions, making it a more suitable model for self-propelled motion. In this study, we focus exclusively on the regime $\lambda > 0$.

It is important here to note a subtle point that is sometimes overlooked: engineered asymmetry in the shape of active particles is not always required for locomotion. Michelin et al. \cite{Michelin} demonstrated that isotropic active particles can undergo a spontaneous symmetry-breaking mechanism, where the resting state becomes unstable to small perturbations and a static-to-dynamic phase transition follows. Studying isotropic active particles, such as self-propelled swimming droplets \cite{Michelin, swimmingdroplet1, swimmingdroplet2, swimmingdroplet3} or disk-shaped camphor grains \cite{Koyano1,Circularchannel,CamphorGrain1dbox,Koyano2,twocamphordisks}, therefore provides a framework for distinguishing self-propulsion mechanisms that arise from intrinsic dynamical instabilities from those that are  imposed by geometric or chemical asymmetry. Making this distinction is useful for clarifying the physical origin of emergent motility in active matter systems.

 Clearly, in both biological and engineered chemical systems, the dynamics inherently involves a form of memory, as active agents continuously respond to the evolving chemical trails they leave behind. The dynamics are thus strongly influenced by the particle’s local environment, including external fields and nearby boundaries \cite{KovacsMemoryEffect, Memory2}. At the level of individual particles, such dynamics break time-reversal symmetry \cite{TimeReversalSymmetry1, TimeReversalSymmetry2}, a hallmark of active systems. This irreversibility is known to give rise to a wide range of complex collective phenomena when many self propelled particles interact, such as motility-induced phase separation (MIPS) \cite{MIPS} and vortex formation \cite{CollectiveVortices}. In addition to these insights at the collective scale, a number of studies have investigated the dynamics of individual active particles, either in free space \cite{Grima, Sengupta, Kranzants, Taktikos} or under confinement \cite{Browniancircleswimmer, Koyano1, Circularchannel, HarmonicTrap, dauchotdemery}. However, many existing descriptions rely either on phenomenological modeling approaches, such as run-and-tumble dynamics or active Brownian particle models \cite{dauchotdemery, runandtumble1, Basu2020, ABP2d}, or on time-derivative expansions that are valid only in the limit of rapidly relaxing fields \cite{Koyano1,Koyano2}.

In this work, we investigate the question of how a single self-propelled particle behaves under confinement. Building on  previous work of a negative self-phoretic particle confined in a one-dimensional harmonic potential \cite{HarmonicTrap}, we modify the model to study the effect of geometric confinement in a one-dimensional channel. We demonstrate that the system exhibits a phase transition from a static to an oscillatory state when the phoretic mobility exceeds a critical value, confirming experimental observations \cite{CamphorGrain1dbox, Koyano1}. The resulting model is purely deterministic and is equivalent to that proposed by Koyano et al. \cite{Koyano1} for an isotropic camphor grain confined in a one-dimensional water channel.

In \cite{Koyano1}, the proposed model successfully captured the qualitative features of the experimental observations reported in the same work. The theoretical analysis in \cite{Koyano1} was based on an expansion of the exact equations in the amplitude of motion, while retaining only a finite number of time derivatives. The truncation of higher-order time derivatives, however, prevents a full characterization of the phase diagram separating passive and active regimes. Here, we perform a perturbative analysis up to cubic order in the particle amplitude, which yields the exact phase diagram and provides an excellent quantitative description of the dynamics for amplitudes as large as half the system size. In addition, we investigate the regime in which the oscillation amplitude approaches the system size and determine how close the particle comes to the system boundaries as a function of the phoretic coupling, $\lambda$. Finally, we present an approximation scheme that determines the particle velocity, $V(X_t)$, as a function of its position, $X_t$, along the periodic orbit established in the long-time limit.

This paper is organized as follows: In Sec.~\ref{sec:themodelsystem}, a mathematical model for an active particle in a one-dimensional channel is introduced. The governing equations of motion are expanded over a Fourier series and analytically reduced to a single nonlinear and nonlocal evolution equation for the particle position, $X_t$. In Sec.~\ref{sec:stability}, a stability analysis is performed up to cubic order in the particle position. The linear part of this analysis reveals an oscillatory instability (Hopf bifurcation) occurring at a critical value $\lambda_c$, which can be computed exactly. The results of the stability analysis are presented and discussed, including the identification of the phase diagram of the system and the derivation of  formulas for both the amplitude and frequency of weakly nonlinear oscillations just above the bifurcation threshold $(\lambda \gtrsim\lambda_c)$. Close to the transition, the cubic order analysis allows us to predict the amplitude and frequency of the oscillations for $\lambda > \lambda_c$. In principle, this analysis is expected to be valid only for small oscillation amplitudes; however, comparison with subsequent numerical simulations shows remarkable agreement even for amplitudes exceeding a quarter of the total channel length ($A > \frac{L}{2}$).

In Sec.~\ref{sec:largelambdalimit}, the equations of motion are re-analysed in the regime of large phoretic coupling $(\lambda \gg \lambda_c)$. We present a new analytical treatment based on the observation that, at late times and for large values of $\lambda$, the particle dynamics become periodic and, sufficiently far from the channel boundaries, approach those of a free particle with constant velocity. Under the assumption that the translational velocity varies slowly in the bulk of the channel, we neglect time-derivative terms in the equations of motion, reducing the problem to a self-consistent algebraic equation for the velocity as a function of position, $V_t(X_t)$. Numerically analysing this equation for $V_t>0$ and increasing $X_t$ reveals a critical position, $X_c$, beyond which real, positive velocity solutions cease to exist. We interpret $X_c$ as the location in the channel where the particle reverses direction and is reflected by the boundary. In Sec.~\ref{sec:numerics}, we numerically investigate the model system and compare the numerical findings with the analytical predictions. The physical implications of the system’s behaviour across the various examined regimes are discussed in Sec.~\ref{sec:discussion}, and concluding remarks are provided in Sec.~\ref{sec:conclusion}.

\section{The model system} 
\label{sec:themodelsystem}
\subsection{Equations of motion}
We consider a diffusiophoretic particle that emits a chemical field of concentration $c(x,t)$ into its local environment at a constant release rate $s$. The particle is bounded on a one-dimensional domain of size $2L$ with positional coordinate $x \in [-L,L]$. The mathematical model for this system then consists of the evolution equations for the particle position, ${X}_t \equiv {X}(t)$, and the chemical field $c(x,t)$. The chemical field obeys the diffusion equation with a diffusion constant $D$, an evaporation rate $\mu$, and a source term 
\begin{equation} 
\frac{\partial c(x,t)}{\partial t} = Dc''(x,t) -\mu \ c(x,t)+ s \ \delta(x- {{X}_t} ), \label{1}
\end{equation}
where the prime notation used here denotes a spatial derivative: $c'(x,t) := \frac{\partial c(x,t)}{\partial x}$. In this model, the particle is taken to be  point-like and thus the corresponding source term is a Dirac delta function  at the particle position, $x= {X}_t$. The particle position obeys equation \eqref{0}. In the rest of this paper we set, without loss of generality,  $s =1$. The chemical concentration field, $c(x,t)$, is also subject to the Neumann boundary condition
\begin{equation} \label{Bcs}
    c'(x=\pm L,t)= 0 ,
\end{equation}
enforcing that there is no diffusive flux of the chemical field across the domain boundaries. Reflection of the particle itself near the boundaries is induced by the field reflection, so that the particle cannot escape the channel. Equations \eqref{0}, \eqref{1} and \eqref{Bcs} then make up the governing equations of motion for the system.

\subsection{Fourier series expansion}
By  symmetry, it is clear that a particle initially placed in the center of the system, ${X}_0=0$, feels identical phoretic forces on the left and right and thus it appears that the state ${X}_t =0$ is an equilibrium state. However, this stationary solution can  be unstable to small perturbations and transform into a state  with non-trivial dynamics. In this subsection, we reduce equations~\eqref{0}–\eqref{Bcs} to a non-linear evolution equation for $X_t$ that is explicitly non-local in time, reflecting the memory induced by the field dynamics. We begin by expanding the phoretic field, $c(x,t)$, as a Fourier series  
\begin{equation}
    c(x,t) = \sum_{n=0}^{\infty} a_n(t) \phi_n (x), \label{37}
\end{equation}
where the field mode coefficients, $a_n(t)$, have only temporal dependence. The spatial eigenmodes, $\phi_n(x)$, satisfy the Neumann boundary condition, $\phi_n'(\pm L) = 0$, and obey the eigenvalue equation 
\begin{equation}
    H \phi_n(x) = \epsilon_n \phi_n(x), \label{38}
\end{equation}
with eigenvalues, $\epsilon_n$, and linear spatial operator $H \equiv - D \frac{d^2}{d x^2}+ \mu $. The normalised eigenfunction is
\begin{equation}
    \phi_n(x) = \sqrt{\frac{1}{L}} \cos{(\frac{n \pi x}{2 L} + \frac{n \pi}{2})}, \label{40}
\end{equation}
with corresponding eigenvalues $\epsilon_n = \frac{D n^2 \pi^2}{4 L^2} + \mu.$ The point source delta function term may then be decomposed in terms of the spatial eigenfunctions, $$\delta(x-{X}_t) = \sum_n \phi_n(x) \phi_n({X}_t),$$ such that equation \eqref{1} reduces to the following ordinary differential equation
\begin{equation} \label{anODE}
\frac{d{a_n}(t)}{dt}+ \epsilon_n {a_n}(t) =   \phi_n({X}_t).
\end{equation}
Substituting the solution $a_n(t)$ in \eqref{0} we find the following nonlinear and nonlocal equation for the process $X_t$ 
\begin{equation}
    \frac{d{X}_t}{dt}  = - \lambda   \sum_{n=1}^\infty \phi_n'({X}_t)\Big[\frac{d}{dt}+ \epsilon_n \Big]^{-1} \phi_n({X}_t) + \xi(t), \label{double}
\end{equation}
where the $n=0$ mode is excluded as it does not contribute to the dynamics ($\phi_0' ({X}_t) = 0$).
We have also introduced an additional term $\xi(t)$, which represents an arbitrary external force, in order to compute the system's response function. While many models in the literature treat this term as a Gaussian white noise to model the effect of thermal Brownian motion \cite{reviewSPP, Stenhammar, MIPS}, we do not include any stochasticity here. The dynamics therefore remain fully deterministic, governed solely by the particle's initial conditions.

\section{Stability analysis}\label{sec:stability}
In this section, we analyze the stability of the stationary solution $X_t=0$ by expanding equation \eqref{double} perturbatively for small displacements ($X_t\ll1$). At linear order, this procedure leads to an integro-differential equation with a memory kernel, and the corresponding linear response function reveals the location of a static-to-oscillatory phase transition. Extending the perturbation expansion to cubic order in $X_t$ yields the frequency and amplitude of the oscillations relatively close to the transition (however, the frequency for very small oscillations at the onset of the transition is also determined by the linear analysis). 

\subsection{Derivation of linear and cubic order response functions}
We assume that, at all times, the particle remains sufficiently close to its initial position at the mid-channel ($x=0$), such that $|{X}_t| \ll 1$. Taylor expanding equation \eqref{double} to cubic order in ${X}_t$, we have
\begin{equation}
        \frac{d{X}_t}{dt}  \approx - \lambda  \sum_{n=1}^\infty \frac{1}{\epsilon_n} \Big[\phi_n'(0) + \phi_n''(0) {X}_t + \phi_n'''(0)\frac{{X}_t^2}{2}  + \phi_n^{(4)}(0) \frac{{X}_t^3}{6} \Big] f(t)  +  \xi(t), \label{cubic}
\end{equation}
where \begin{equation}
    f(t) \equiv \Big[1+\frac{1}{\epsilon_n}\frac{d}{dt}\Big]^{-1} \Big[\phi_n(0) + \phi_n'(0){X}_t + \phi_n''(0)\frac{{X}_t^2}{2} + \phi_n'''(0)\frac{{X}_t^3}{6} \Big ].
\end{equation}
For convenience, we define the function $g(t) \equiv f(t) - \phi_n(0) $, which can be written to cubic order as $g(t) = \sum_{i=1}^3 g_i(t)$, with
\begin{equation}
    g_i(t) = \Big [ 1 + \frac{1}{\epsilon_n}\frac{d}{dt}  \Big]^{-1}  \frac{\phi_n^i (0) ({X}_t) ^i}{i!}.
\end{equation}
Equation \eqref{cubic} then reduces to

\begin{equation}
    \begin{aligned}
  \frac{d{X}_t}{dt}  \approx - \lambda  \sum_{n} \frac{1}{\epsilon_n} \bigg[ &  \phi_n'(0) \big[g_1(t) + g_3(t) \big] + {X}_t \phi_n''(0) \big[g_2(t) + \phi_n(0) \big] \\& + \frac{ {X}_t^2}{2} \phi_n'''(0) g_1(t) + \frac{{X}_t^3}{6} \phi_n^4(0) \phi_n(0)\bigg] + \xi(t),
    \end{aligned}\label{gcubiceqn}
    \end{equation}
where we have kept only terms up to order ${X}_t^3$. We note that all even-order terms in ${X}_t$, being proportional to $\cos{(\frac{n \pi}{2}})\sin{(\frac{n \pi}{2}})$, vanish identically for all $n$. The survival of only odd-order terms in ${X}_t$ is a  direct consequence of the centrosymmetry of the system. At linear order, we have the integro-differential equation
\begin{equation}
     \frac{d{X}_t}{dt}  + \lambda \bigg( X_t\sum_{n}\frac{\phi_n(0)\phi_n''(0)}{\epsilon_n} + \sum_{n}\big[\phi_n'(0)\big]^2 e^{-\epsilon_n t} \int_{-\infty}^{t}  X_s e^{\epsilon_n s} ds  \bigg) = \xi(t),
\end{equation}
which shows the explicit dependence on the entire previous trajectory of the particle. Now if we apply an external force $\xi(t)=\exp(i\omega t)$, we obtain a solution of the form $X(t) = R (\omega)\exp(i\omega t)$, where $R(\omega)$ is the linear response function
\begin{equation}
    \big [  R (\omega) \big ] ^{-1} = i \omega + \lambda  \bigg (\sum_{n=1}^\infty \frac{\phi_n''(0) \phi_n(0) }{\epsilon_n} +  \sum_{n=1}^\infty \frac{[\phi_n'(0)]^2}{\epsilon_n + i \omega} \bigg ). 
\end{equation}
Divergences of the response function at real frequencies signal instabilities of the system to perturbations at those same frequencies. Mathematically, these unstable frequencies correspond to zeros of the inverse response function, $   \big [  R (\omega) \big ] ^{-1}$.

To determine the cubic order terms, we return to equation \eqref{gcubiceqn} and look for a periodic solution of the form $X(t)= A\big(\exp(i\omega t)+\exp(-i\omega t)\big)$ and set $\xi(t)=0$.  Keeping only terms up to order $A^3$, we find 
\begin{equation}
         { \big[ R (\omega)\big]} ^{-1} + \frac{\lambda {A}^2}{16}  \sum_{n=1}^\infty \Big [{ \frac{4 [\phi_n''(0)]^2}{\epsilon_n} +    \frac{2 [\phi_n''(0)]^2}{\epsilon_n + 2i\omega}  + \frac{2 \phi_n'(0)\phi_n'''(0)}{\epsilon_n - i \omega} + \frac{6 \phi_n'(0)\phi_n'''(0)}{\epsilon_n + i \omega} + \frac{2 \phi_n(0)\phi_n^{(4)}(0)}{\epsilon_n }  }\Big] =0,
\end{equation}
or the passive solution $A=0$. As before, we have matched only the terms proportional to $e^{i\omega t}$. Although cubic nonlinearities also generate components at $e^{\pm 3 i\omega t}$, results from numerical simulations in Sec.~\ref{sec:numerics} reveal that these higher harmonics are small near the phase transition, and the dynamics are dominated by the single mode at fundamental frequency $\omega$. We therefore omit a detailed calculation of the anharmonic contributions, referring instead to \cite{HarmonicTrap}, where the method is presented for an active particle in a harmonic trap.

To avoid evaluating the summations over $n$ explicitly, it is useful to consider the Green's function
 \begin{equation}
   {G}(x,y, \omega)  = \sum_{n=1}^\infty \frac{\phi_n(x) \phi_n(y)}{(\epsilon_n + i \omega)}, \label{greens}
\end{equation} which satisfies $$[H + i \omega] {G}(x,y,\omega) = \delta(x-y),$$ as well as the Neumann boundary condition $\frac{\partial G}{\partial x}(x = \pm L,y,\omega)=0$. The response function, $R(\omega)$ and the equation for the amplitude $A$, can then be written entirely in terms of the Green's function \eqref{greens}, as 
\begin{equation}
    {[ R (\omega) ]}^{-1} = i\omega + \lambda \bigg[ \frac{\partial ^2 G}{\partial x^2} {(0,0,0)} + \frac{\partial ^2 G}{\partial x \partial y} {(0,0,\omega)} \bigg ],\label{linearrespG}
\end{equation}
and 
\begin{equation}
\begin{aligned}
       0 = {[ R (\omega)}] ^{-1} + \frac{\lambda  {A}^2}{16} & \bigg [ 4 \frac{\partial ^4 G}{\partial x^2 \partial y^2} {(0,0,0)} + 2  \frac{\partial ^4 G}{\partial x^2 \partial y^2} {(0,0,2\omega)} 
    + 2  \frac{\partial ^4 G}{\partial x \partial y^3} {(0,0,-\omega)} \\ &
    + 6  \frac{\partial ^4 G}{\partial x \partial y^3} {(0,0,\omega)} 
    + 2 \frac{\partial ^4 G}{\partial y^4} {(0,0,0)}
    \bigg].\label{cubicresponse}
\end{aligned}
\end{equation}
The Green's function solution is derived by considering the homogeneous equation
$$ \frac{\partial ^2 }{\partial x^2} G(x,y,\omega) - m^2 G(x,y,\omega) =0,$$ where we define $ m^2 \equiv \frac{\mu + i\omega}{D}$ with $\Re({m^2}) > 0$. It might appear that evaluating the derivatives of the Green's functions at coinciding points is delicate as they have derivative discontinuities. However one can verify that the derivatives of delta functions arising in Eq. (\ref{cubicresponse}) vanish.
Then the general solution that satisfies the Neumann boundary condition is
\begin{equation}
G = \begin{cases} 
      B \cosh(m(x+L)) &; x < y \\
      \tilde B \cosh(m(x-L)) &; x >y 
   \end{cases}. \label{13}
\end{equation}
The constants $B$ and $\tilde B$ are set by enforcing continuity of $G$ at $x=y$ and the jump condition $ \frac{d G(x=y^-)}{dx}-\frac{d G(x=y^+)}{dx}= \frac{1}{D}$, which leads to the complete Green's function solution
\begin{equation}
G(x,y, \omega) = \begin{cases}
    \frac{\cosh(m(y-L))\cosh(m(x+L)}{m D \sinh(2 mL)} &; x<y \\
     \frac{\cosh(m(y+L))\cosh(m(x-L)}{m D \sinh(2 mL)} &;  x>y
\end{cases}. \label{greensoln}
\end{equation}
Evaluating \eqref{linearrespG} with \eqref{greensoln} then gives the inverse linear response function 
\begin{equation}
    \big [ R(\omega) \big]^{-1} = 
     i \omega + \frac{\lambda  }{2 D} \bigg(  \sqrt{\frac{\mu}{D}} \coth{\Big( L \sqrt{\frac{\mu}{D}}\Big)} -   \sqrt{\frac{\mu + i \omega}{D}} \tanh{\Big(L \sqrt{\frac{\mu+i\omega}{D}}\Big)}  \bigg ).
     \label{linearrespfn1}
\end{equation} 
Evaluating \eqref{cubicresponse} similarly gives the amplitude equation
\begin{equation}
     0= {[ R (\omega)]} ^{-1} + \frac{\lambda  {A}^2}{16 D} 
 h(\omega, \mu, L ,D), \label{cubicresp2}
\end{equation}
where we define
\begin{equation}
\begin{aligned}
    h(\omega, \mu, L ,D) \equiv \ & 3 \sqrt{\frac{\mu}{D}}^3 \coth{\Big(L \sqrt{\frac{\mu}{D}}\Big)} + \sqrt{\frac{\mu + 2 i \omega}{D}}^3 \coth{\Big(L \sqrt{\frac{\mu + 2 i \omega}{D}}\Big)} \\ & -  \sqrt{\frac{\mu - i \omega}{D}}^3 \tanh{\Big(L \sqrt{\frac{\mu - i \omega}{D}}\Big)} - 3 \sqrt{\frac{\mu + i \omega}{D}}^3 \tanh{\Big(L \sqrt{\frac{\mu + i \omega}{D}}\Big)} .
\end{aligned}
\end{equation}

\subsection{Free particle limit $(L \rightarrow \infty)$}\label{sec:freeparticle}
We note here that it is interesting to consider the inverse linear response function \eqref{linearrespfn1} in the limit of an infinitely large channel, $L \rightarrow \infty$. This recovers the explicit result of \cite{HarmonicTrap} for a free, point-source particle in one-dimension, in the absence of any confining potential: 
\begin{equation}
    \lim_{\substack{L \to \infty}}   \Big[ R(\omega ) \Big]^{-1} = i \omega + \frac{\lambda }{2 D^\frac{3}{2}} \Big [\sqrt{\mu} - \sqrt{\mu + i \omega} \Big]. \label{30}
    \end{equation}
In this regime, the particle undergoes a spontaneous-symmetry breaking transition from a rest phase to a constant velocity phase. Solving for the poles of \eqref{30} gives the location of the phase transition at a critical parameter value $\lambda_c = 4 D^{\frac{3}{2}} \sqrt{\mu}$. This recovers exactly the stability criterion reported by Grima \cite{Grimaresult} for a free particle with zero time delay in the source term, which is equivalent to our model for a free particle with zero particle size. The velocity for a free point-like particle on an infinite one-dimensional domain can be calculated straightforwardly by assuming a steady-state constant velocity solution, $\frac{dX_t}{dt}=v$, to equations \eqref{0} and \eqref{1}. Performing a Galilean transformation from the laboratory frame to the rest frame of the particle, we make the change of variable $y := x - X_0 -vt$. Equation \eqref{1} then becomes $$-D \ c''(y) - v  \ c'(y) +\mu \ c(y) = \delta(y),$$ which has the Green's function solution
\begin{equation}\label{cy}
    c(y) = \frac{1}{\sqrt{v^2 +4D\mu}} \exp{\bigg(- \frac{vy + |y|\sqrt{v^2+4D\mu}}{2D} \bigg)}. 
\end{equation}
Clearly, from equation \eqref{cy} we can see that the derivative $\frac{dc}{dy} \equiv c'(y)$ is not continuous at $y=0$ (or, in the lab frame: the particle position $x=X_0+vt$). From Fourier analysis, or by regularising the delta function, the value of $c'(0)$ is given by its average over the left and right of the particle position. Using $c'(0) =\frac{1}{2}\big(c'(0^+) + c'(0^-)\big)$ to evaluate equation \eqref{0} gives
\begin{equation}\label{vroot}
    v= \frac{\lambda v}{2D\sqrt{v^2 +4 D\mu}},
\end{equation}
which has solutions at $v=0$ or $v=\pm\frac{\sqrt{\lambda^2-16D^3\mu}}{2D}$. The zero and non-zero velocity solutions are the solutions stable to perturbation for $\lambda<\lambda_c$ and $\lambda>\lambda_c$, respectively.

\subsection{Location of the phase transition} \label{sec:linearanalysis}
By causality, the time-domain response function must satisfy $R(t)=0$ for $t<0$. Using the Fourier transform convention\[R(t)=\frac{1}{2\pi}\int_{-\infty}^{\infty} R(\omega)\, e^{i\omega t}\, d\omega,\] and
evaluating $R(t)$ via Cauchy’s Residue Theorem requires that $R(\omega)$ have no poles in the lower half of the complex $\omega$-plane. The stationary solution, $X_t = 0$, then becomes unstable when $R(\omega)$ develops a pole on the real axis. If such a pole appears and crosses the real axis at some critical frequency $\omega_c$ (with a corresponding pole simultaneously crossing at $-\omega_c$), a bifurcation-type phase transition occurs from a static state to an active one. In this case, the instability of the fixed point $X_t=0$ implies the emergence of a stable limit cycle, corresponding physically to oscillatory motion of the particle about the system center. Exactly at the transition, the particle’s oscillation frequency is $\omega_c$, and the associated critical parameter values are $\{\mu_c \equiv \mu(\omega_c), \lambda_c \equiv \lambda(\omega_c)  \}$. These critical values can be identified from solving the complex equation $\big [ R(\omega_c ) \big]^{-1} = 0$. Switching now to dimensionless variables, we find that the critical point is determined by the following equations
\begin{equation}\label{realpt}
    \operatorname{Re} \Big [\sqrt{\mu_c^* +i \omega_c^*} \tanh{(\sqrt{\mu_c^*+ i \omega_c^*})} \Big]=\sqrt{\mu_c^*}\coth{(\sqrt{\mu_c^*})} 
\end{equation}
\begin{equation}\label{imagpt}
    \operatorname{Im} \Big [ \sqrt{\mu_c^*+i \omega_c^*} \tanh{\sqrt{\mu_c^*+i \omega_c^*}}\Big] = \frac{2\omega_c^*}{\lambda_c^*},
\end{equation}
where we have introduced the dimensionless parameters $\mu^* \equiv\frac{\mu L^2}{D}$, $\omega^* \equiv \frac{\omega L^2}{D}$ and $\lambda^* \equiv \frac{\lambda L }{D^2}$. 
These critical equations give the line of critical values of $\mu_c^*$ and $\lambda_c^*$,  as well as the critical value of the frequency at the transition, $\omega^*_c$. In fact, they define a phase boundary curve in the $\{\mu^*, \lambda^*\}$ plane space that is parametrized by $\omega_c^*$.

\begin{figure}[htb]
    \centering
    \captionsetup[subfigure]{labelformat=empty} 
    \begin{subfigure}{0.45 \textwidth}  
        \centering
        \includegraphics[width=\linewidth]{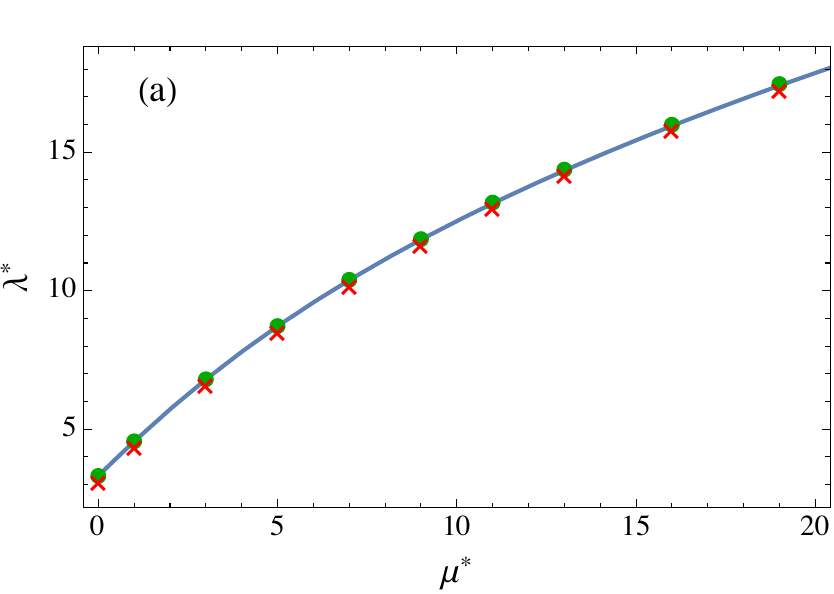}
    \end{subfigure}
    \hfill
    \begin{subfigure}{0.45 \textwidth}
        \centering
        \includegraphics[width=\linewidth]{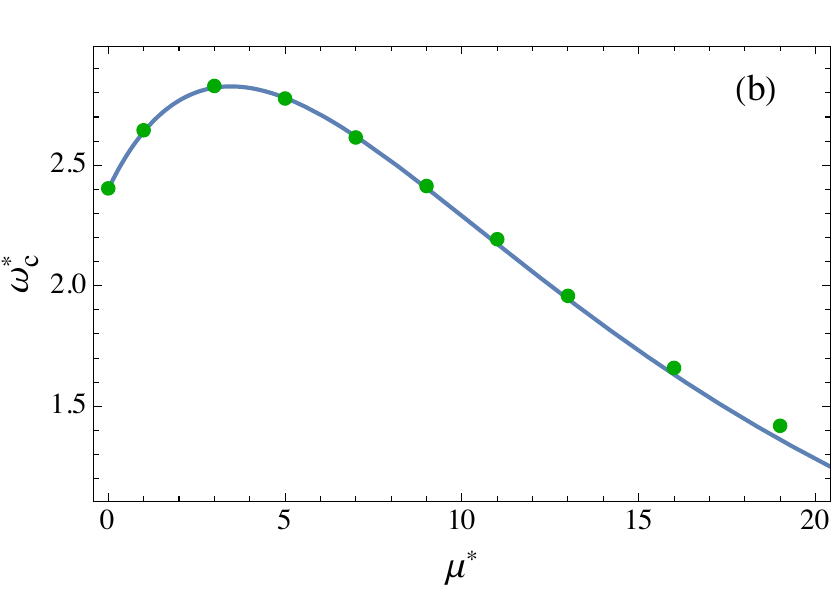}
    \end{subfigure}
    \caption{\small \textbf{(a)} Stability boundary between oscillatory (above) and static (below) phases in the $ \{ \mu^*,\lambda^* \}$ plane. The blue line shows the analytical prediction; green points denote the numerically determined phase boundary, and red crosses indicate simulations that remain static at long times. \textbf{(b)} Critical frequency at the transition, $\omega_c^*$, versus absorption $\mu^*$. The blue line is the analytical prediction and green points are numerical results.}
    \label{fig:phasediagram}
\end{figure}

The blue curve plotted in Fig.~\ref{fig:phasediagram}\textcolor{blue}{(a)} shows the phase boundary $\{\mu_c^*, \lambda_c^* \}$ obtained by numerically solving the  equations \eqref{realpt} and \eqref{imagpt}. Fig.~\ref{fig:phasediagram}\textcolor{blue}{(b)} plots the corresponding values for the critical frequency, $\omega_c^*$, at the onset of the phase transition as a function of the absorption, $\mu^*$. From the definition of the dimensionless parameter $\lambda^*$, we find that the critical phoretic mobility obeys the scaling $\lambda_c \propto \frac{D^2}{L}$. This implies that the active oscillatory phase is favored in long channels and for small field diffusion coefficients. This result is consistent with the experimental observations of Koyano et al. \cite{Koyano1}, who reported that for short one-dimensional water channels a camphor grain remains stationary due to camphor molecules rapidly saturating the water surface and failing to generate an adequate driving force to set the particle in motion. As a result, the grain settles at the mid-channel position where the residual forces balance. However, when the channel length is increased the saturation process slows, enabling the camphor grain to acquire sufficient driving force to initiate motion and transition into the active, oscillating phase. 

From Fig.~\ref{fig:phasediagram}\textcolor{blue}{(b)} we also see that as $\mu^*$ increases, $\omega^*_c$ rises to a single maximum at $\mu^* \approx 3.45$ and then decays to zero as $\mu^* \to \infty$. In contrast, $\lambda^*$ grows monotonically and follows a square-root law for large $\mu^*$, scaling asymptotically like $\lambda^* \sim 4\sqrt{\mu^*}$. This behaviour can be physically understood, as for a larger evaporation rate the trail of chemical field vanishes quicker and the memory effect becomes weaker, therefore requiring a larger particle-field coupling to obtain an instability. 

\subsection{Amplitude equation}
When a real solution exists, the amplitude equation can be written in terms of dimensionless variables as
\begin{equation}\label{amplitudeeqn}
   A^*  = \sqrt{\frac{- 16  \big[R^*(\omega^*)\big]^{-1} }{\lambda^*  \ h^*(\omega^*, \mu^*)}},
\end{equation}
where \begin{equation}
    \big [ R^*(\omega^*) \big]^{-1} = 
     i \omega^* + \frac{\lambda^*  }{2} \bigg(  \sqrt{\mu^*} \coth{ \sqrt{\mu^*}} -   \sqrt{\mu^* + i \omega^*} \tanh{\sqrt{\mu^*+i\omega^*}}  \bigg ),
     \label{linearrespfn}
\end{equation} 
and \begin{equation}
\begin{aligned}
    h^*(\omega^*, \mu^*) \equiv \ & 3 \sqrt{{\mu^*}}^3 \coth{\sqrt{\mu^*}} + \sqrt{\mu^* + 2 i \omega^*}^3 \coth{ \sqrt{\mu^* + 2 i \omega^*}} \\ & -  \sqrt{\mu^* - i \omega^*}^3 \tanh{ \sqrt{{\mu^* - i \omega^*}}} - 3 \sqrt{\mu^* + i \omega^*}^3 \tanh{ \sqrt{{\mu^* + i \omega^*}}} .
\end{aligned}
\end{equation}
The amplitude must be purely real, whereas the right hand side of \eqref{amplitudeeqn} is complex and therefore gives two equations. Since the imaginary part of $A^*$ must be zero, the equation ${\Im}[A^*]=0$ can be used to determine the oscillation frequency $\omega^*$, and then the amplitude can be determined from the real part.

\section{Large $\lambda$ limit}\label{sec:largelambdalimit}
In the regime of large phoretic mobility, $\lambda \gg \lambda_c$, the particle’s oscillation amplitude grows to order $L$, the channel half-length. Consequently, the stability analysis developed in Sec. \ref{sec:stability} no longer applies. We therefore return to the dimensional governing equations, \eqref{anODE} and \eqref{double}, and write the field modes $a_n(t)$ as
\begin{equation}\label{ansatz}
    a_n(t) = u_n(t) \phi_n(X_t) + w_n(t) \phi_n'(X_t),
\end{equation}
where we have introduced the coefficients $u_n(t)$ and $w_n(t)$. Substituting \eqref{ansatz} into \eqref{anODE} and equating coefficients of $\phi_n(X_t)$ and $\phi_n'(X_t)$ gives the following coupled equations
\begin{equation}\label{unwn1}
   \dot{u}_n(t) +  \epsilon_n u_n(t) + \frac{w_n(t) V_t (\mu - \epsilon_n)}{D} =1,
\end{equation}
\begin{equation}\label{unwn2}
  \dot{w}_n(t) +   \epsilon_n w_n(t) + u_n(t) V_t =0,
\end{equation}
where we have also used the fact that $\phi_n''(X_t) = (\frac{\mu-\epsilon_n}{D}) \phi_n(X_t)$. From the results of section \ref{sec:freeparticle}, we know that the particle's behaviour in the region far from the channel boundaries mimics that of a free particle in an infinite system. Thus, we assume that the particle's translational velocity varies slowly in the bulk of the channel, such that we may neglect the explicit time derivative terms $ \dot{u}_n(t) =  \dot{w}_n(t)=0$. Solving the resulting algebraic equations, we find
\begin{equation}
    u_n(t) = \frac{D\epsilon_n}{D\epsilon_n^2 + V_t^2(\epsilon_n -\mu)},
\end{equation}
\begin{equation}
    w_n(t) =  \frac{-D V_t}{D\epsilon_n^2 +V_t^2(\epsilon_n-\mu)}.
\end{equation}
Substituting these results back into the noiseless equation \eqref{double} (with $\xi(t)=0$), we find that the particle velocity, $V_t := \frac{dX_t}{dt}$, obeys 
\begin{equation}\label{Vt}
    V_t = - \lambda \sum_n \frac{D\epsilon_n \phi_n(X_t) \phi_n'(X_t) -D V_t [\phi_n'(X_t)]^2}{D\epsilon_n^2 + V_t^2(\epsilon_n -\mu)} . 
\end{equation}
The quadratic expression in the denominator of \eqref{Vt} has roots $r_{\pm} = \frac{-V_t^2 \pm \sqrt{V_t^4 + 4 \mu D V_t^2}}{2 D}$, allowing us to rewrite using partial fractions. We find
\begin{equation}\label{partialfraceqn}
    V_t = - \frac{\lambda  }{(r_+ - r_-)} \sum_n  \Big [\frac{r_+ \phi_n(X_t) \phi_n'(X_t)}{\epsilon_n - r_+} -\frac{r_- \phi_n(X_t) \phi_n'(X_t)}{\epsilon_n - r_-}  -\frac{V_t [\phi_n'(X_t)]^2}{\epsilon_n- r_+} + \frac{V_t[\phi_n'(X_t)]^2}{\epsilon_n- r_-} \Big].
\end{equation}
Now each term of \eqref{partialfraceqn} can be defined in terms of the Green's function \eqref{greens}, $G(x,y,\omega = i\alpha)$, where $\alpha$ is a real constant
\begin{equation}\label{vtgreenseqn}
     V_t = - \frac{\lambda  }{(r_+ - r_-)} \bigg[r_+ \frac{\partial G}{\partial x} (X_t,X_t,i r_+) - r_- \frac{\partial G }{\partial x}(X_t,X_t,i r_-) - V_t\frac{\partial^2 G}{\partial x \partial y} (X_t,X_t, ir_+) + V_t\frac{\partial^2 G}{\partial x \partial y} (X_t,X_t,i r_-)
     \bigg].
\end{equation}
Given the known Green's function solution \eqref{greensoln}, clearly the derivative terms $\frac{\partial G}{\partial x}(x,y,i\alpha)$ are discontinuous at $x=y$. Then, following the same argument as used for equation \eqref{vroot}, the first two terms of \eqref{vtgreenseqn} must be interpreted as the average of $\frac{\partial G}{\partial x}$ to the left and right of $x=y$. The second derivative terms are continuous either side of $x=y$, and straightforward to calculate. The resulting equation is given by
\begin{equation}\label{j}
    V_t  = \frac{\lambda}{D}  \ \theta(X_t,V_t, \mu,D, L),\end{equation}
where we define the function
\begin{equation}\label{largeeV}
\begin{aligned}
       \theta(X_t,V_t, \mu,D, L) := & \frac{1}{4 S} \bigg [(V_t - S) \Big \{\coth{\big(\frac{L}{D}(S-V_t)\big) - e^{\frac{X_t}{D}(V_t-S)} \operatorname{cosech}{\big(\frac{L}{D}(S-V_t)\big)} }\Big\} \\&  +(V_t+S) \Big\{\coth{\big(\frac{L}{D}(S+V_t)\big) - e^{\frac{X_t}{D}(V_t+S)} \operatorname{cosech}{\big(\frac{L}{D}(S+V_t)\big)}} \Big \} \bigg]
       \end{aligned}
\end{equation}
with $S(V_t,\mu,D) \equiv \sqrt{ V_t^2 +4D\mu}$. Evaluating \eqref{j} at $X_t=0$, we find that the mid-channel velocity in the $\lambda \gg \lambda_c$ limit obeys
\begin{multline}\label{largelambdV}
        V_t(X_t=0) = \frac{\lambda  }{ 4 D S}
        \bigg[{(V_t - S)}\tanh{\big(\frac{L}{2D}(S-V_t)\big)} +{(V_t+S)} \tanh{\big(\frac{L}{2D} (S+V_t) \big)} \bigg].
\end{multline}

We note that \eqref{j} is a self-consistent algebraic equation for the translational velocity, $V_t$, and may be analysed using a phase-portrait interpretation, where the velocity is viewed as a function of the particle’s position within the channel, $V_t(X_t)$. In Appendix~(\ref{sec:image}), we recover the same result using an image-charge–type construction. Specifically, we consider an unconfined system augmented by identical “mirror image” particles located outside the physical domain ($|x| > L$). In this framework, we again compute the constant-velocity solution, with the image particles enforcing the Neumann boundary conditions. The resulting solution is numerically identical to that obtained above. Physically, this shows that reflection at the boundary $x=L$ is dominated by the interaction between the particle, as it approaches the boundary ($X_t \rightarrow L$), and its nearest image at $x=2L-X_t$. This approach effectively renormalizes the spatial gradient of the field $c(x,t)$ experienced by a free particle and leads to reflection as the particle and its image interact, thereby reducing the velocity of both.

\begin{figure}[htb]
    \centering
    \captionsetup[subfigure]{labelformat=empty} 
    \begin{subfigure}{0.45 \textwidth}  
        \centering
        \includegraphics[width=\linewidth]{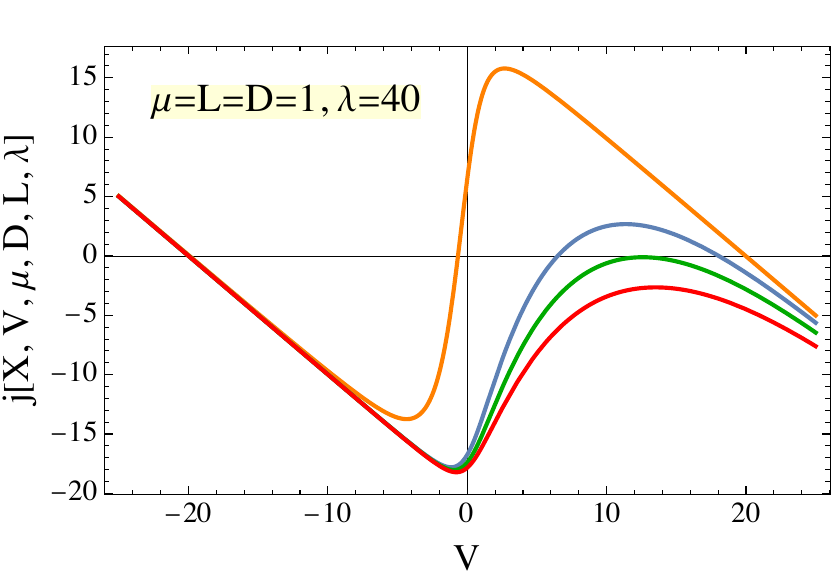}
    \end{subfigure}
    \begin{subfigure}{0.15 \textwidth}
        \centering
        \includegraphics[width=\linewidth]{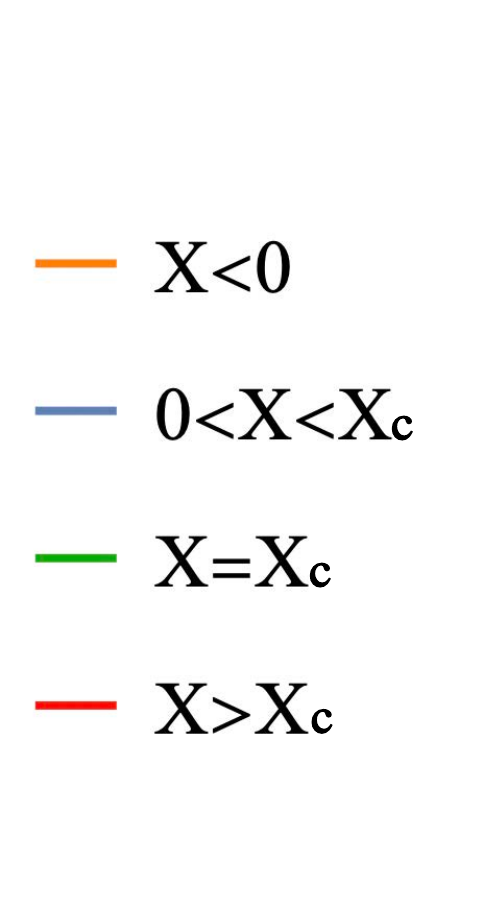}
    \end{subfigure}
    \caption{\small Numerical solutions of $j[X,V, \mu, D, L,\lambda]=0$, with $j := V- \frac{\lambda}{D} \theta(X,V, \mu, D, L)  $. For $\mu, D, L =1$ and $\lambda=40$ the equation admits at most three real roots depending on $X$. A physically invalid root at $v_1=-20$ exists for all $X$. For $0<X<X_c$ there are two positive roots at $v_2,v_3$ with $v_2 < v_3$. Beyond a critical value of $X_c$ no positive roots exist, and for $X<0$ the root $v_2$ becomes negative. Curves shown correspond to $X=-0.5$ (orange), $X= 0.915< X_c$ (blue), $X=X_c=0.9326$ (green) and $X=0.945 >X_c$ (red).}
    \label{fig:3vsolns}
\end{figure}

Numerical analysis of the equation \eqref{j} shows that, for fixed parameter values $\mu, D, L, \lambda$, there exist three real solutions for the velocity $V_t$ when the position $X_t$ is increased from zero: one negative solution $(v_1)$ and two distinct positive solutions $(v_2, v_3$ with $v_3>v_2)$. We consider the phase where  motion of the particle is such that $V_t>0$ (so in the left to right part of the trajectoy); this immediately eliminates the negative solution $v_1$. In Fig.~\ref{fig:3vsolns} we demonstrate graphically that the multiplicity of positive solutions persists only for $X_t>0$ and up to a critical position $X_c$. For positions beyond this critical value ($X_t>X_c$), no positive solution for the velocity exists. We therefore interpret $X_c$ as the position in the channel at which the velocity changes rapidly and the particle undergoes a reversal of direction, effectively being reflected. In other words, $X_c$ provides a theoretical prediction for the amplitude of oscillation deep in the active phase. For $X_t<X_c$, the physically relevant solution for the velocity is $v_3$, the largest positive root of the algebraic equation. This choice is justified because for negative positions ($X_t<0$) $v_2$ becomes negative and so, by continuity, $v_3$ is the only physically meaningful solution throughout the entire region $X_t<X_c$ and 

In the remainder of this section, we estimate the particle’s closest approach to the ends of the channel using the following argument. We make the change of variable $V_t \rightarrow U \lambda$ and expand equation \eqref{j} for $\lambda \gg 1$ and $U>0$. Keeping terms up to order $\mathcal{O}(\lambda^{-1})$, we find 
\begin{equation}\label{resultU}
U \simeq  \frac{1}{2D} \bigg[1-2 e^{-\frac{2U\lambda \delta}{D}}\bigg].
\end{equation}
where $\delta \equiv {L-X_t}$ is the distance of the particle to the boundary. The result~\eqref{resultU} can also be obtained using the image-particle method described in Sec.~\ref{sec:image}. In this approach, as the particle approaches \(x=L\) from the left, its emitted field \(c(x,t)\) is assumed to coincide with that of a free particle moving at constant velocity \(U\lambda\). An identical field is produced by an image particle approaching \(x=L\) from the right, with image position $\bar X_t = 2L - X_t$ and velocity $\bar V_t = -\,U\lambda$. The velocity of the real particle, \(U\lambda\), is then determined from $-\lambda\, c'(X_t),$ where \(c'(X_t)\) is the sum of the fields generated by the particle and its nearest image. A complete treatment, including higher-order effects, would require accounting for field contributions from the infinite hierarchy of image particles.

Differentiating \eqref{resultU} with respect to U gives $e^{-\frac{2U\lambda \delta}{D}} = \frac{D^2}{2\lambda  \delta}$, which can be substituted back into \eqref{resultU} to give $U = \frac{1}{2D}-\frac{D}{2\lambda\delta}$. Combining these to eliminate $U$ then gives
\begin{equation}\label{fk}
   \frac{1}{2} =   ke^{1 - k} := f(k),
\end{equation}
where we define a variable $k:=\frac{\lambda\delta}{D^2}$. For $U>0$, equation \eqref{fk} has only one solution at $k=2.68$. Therefore, we predict that in the large $\lambda$ regime far from the phase transition, the distance of closest approach between the particle and the boundary scales like $\delta \sim \frac{2.68 D^2}{ \lambda}$.

\section{Numerical Simulations}\label{sec:numerics}
\subsection{Method}

The analytical results presented in Secs. \ref{sec:stability} and \ref{sec:largelambdalimit} can be verified with a numerical analysis of our model. The dimensionless numerical equations we  solve are
\begin{equation} \label{deltaX}
    \frac{ dX^*_\tau}{d\tau} = \lambda^*  \sum_{n=1}^N k_n \alpha_n(\tau) \sin{\Big( k_n ({X^*_\tau} + 1) \Big)},
\end{equation}
\begin{equation}\label{deltaan}
    \frac{d \alpha_n(\tau)}{d\tau} + (k_n^2 + \mu^*)\alpha_n(\tau) = \cos{\Big( k_n ({X^*_\tau} + 1) \Big)}, 
\end{equation}
where we have introduced the dimensionless parameters $k_n \equiv \frac{n \pi}{2}$, $X^*_\tau \equiv \frac{X_t}{L}$, $\tau \equiv \frac{t D}{L^2}$, $\alpha_n \equiv \frac{D a_n}{L^{\frac{3}{2}}}$, and reintroduced $\mu^* \equiv\frac{\mu L^2}{D}$ and $\lambda^* \equiv \frac{\lambda L }{D^2}$. The coupled differential equations \eqref{deltaX} and \eqref{deltaan} are integrated numerically using \textit{Wolfram Mathematica}'s \textbf{NDSolve} routine with residual-based simplification and a high timestep limit of $M_{steps}=2\times 10^{5}$ to ensure numerical stability. The equations are formulated in a truncated Fourier basis with $N=1000$ modes inside a one-dimensional channel with positional coordinate $x^*\equiv \frac{x}{L}\in [-1, 1]$, and with the initial conditions $X^*_{\tau=0}=0.1$, and $\alpha_n(0) = 0$ for all $n$.

To investigate the threshold between the active oscillatory and passive rest states, we calculated particle trajectories until they sufficiently converged to a limit-cycle or fixed point. After discarding the transient dynamics, the trajectory is analyzed relative to the mid-channel reference point at $x^*=0$. Zero-crossings of $X^*_\tau $ are detected by sign-change analysis and refined with root finding, providing crossing times from which mean oscillation periods and mean frequencies are calculated. In each half-cycle, the maximal deviation from the reference point is determined, yielding the  oscillation amplitude. The particle velocity is obtained by evaluating the numerical derivative of the particle position, $V^*_\tau :=\frac{d X_\tau^*}{d\tau}$, and is obtained directly from the interpolating polynomials generated by \textbf{NDSolve}. To reduce any high-frequency numerical noise in the computed velocity signal, we applied \textit{Mathematica}'s  \textbf{LowpassFilter} function. A cutoff frequency of $(0.7\times$ Nyquist) was selected to be well above the physical frequencies associated with the modal evolution, ensuring that only non-physical high-frequency artifacts are removed.


\subsection{Results}\label{sec:results}
\begin{figure}[htb]
    \centering
    \captionsetup[subfigure]{labelformat=empty} 
    \begin{subfigure}{0.45 \textwidth}
        \centering
        \includegraphics[width=\linewidth]{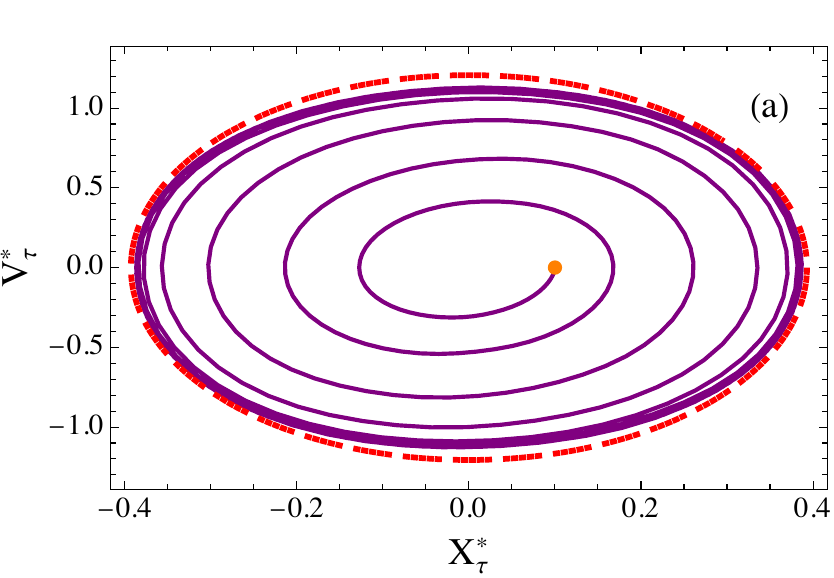}
    \end{subfigure}
    \hfill
    \begin{subfigure}{0.45 \textwidth}
        \centering
        \includegraphics[width=\linewidth]{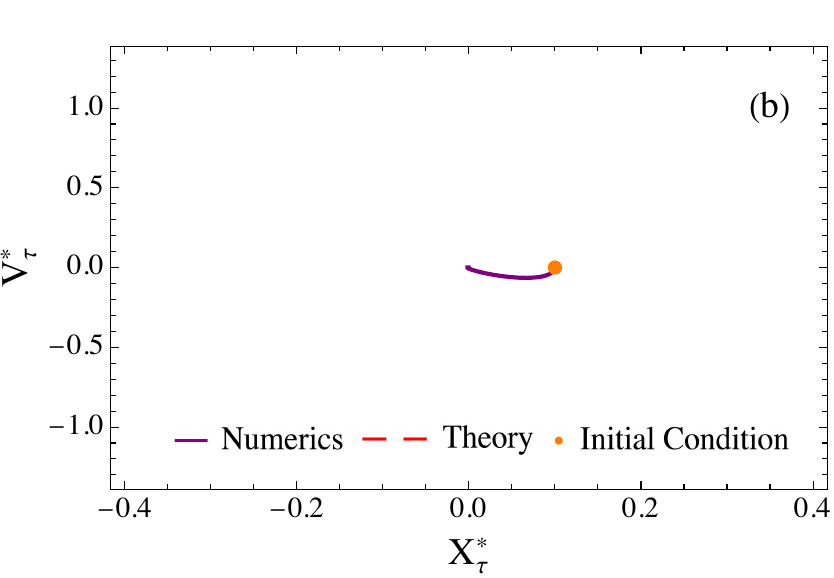}
    \end{subfigure}
    \caption{\small {Representative numerical results as trajectories on the $(X_\tau^*,V_\tau^*)$ plane. The orange circular point indicates the initial position at $(0.1,0)$. \textbf{(a}) For $\mu^*=1, \lambda^*= 5$: Oscillatory motion that settles onto a stable limit cycle after transient growth. Numerics shown by solid purple line and theory by red dashed line. \textbf{(b)} For $\mu^*=1, \lambda^*= 2$: Overdamped oscillations that decay to the rest state at $X_\tau^*=0$.}}
    \label{fig:limitcycle}
\end{figure}

 Representative numerical results from typical simulations are presented in Fig.~\ref{fig:limitcycle} as trajectories on the position-velocity $(X_\tau^* , V_\tau^*)$ phase space. Depending on the values of the  parameters $\mu^*$ and $\lambda^*$, the particle either begins to oscillate with increasing amplitude before converging on a stable limit cycle, as illustrated in Fig.~\ref{fig:limitcycle}\textcolor{blue}{(a)}, or undergoes damped oscillations, converging toward the mid-channel position, as shown in Fig.~\ref{fig:limitcycle}\textcolor{blue}{(b)}. These two distinct dynamical regimes correspond, respectively, to the active and passive phases. The orange circular marker indicates the particle's initial position at $\tau=0$, and the purple solid line represents the simulation trajectory in the plane for increasing $\tau$. At large times in Fig.~\ref{fig:limitcycle}\textcolor{blue}{(a)}, the steady-state trajectory shows excellent agreement with the stability analysis prediction (dashed red line). We note that the predicted trajectory from the cubic order stability analysis assumes perfectly harmonic oscillations close to the phase transition, and thus the analytical prediected trajectory is simply the ellipse defined by $\{A^* \cos{(\omega^*\tau)} , -A^* \omega^* \sin{(\omega^*\tau})\}$, with $A^*$ and $\omega^*$ the solutions of equation \eqref{amplitudeeqn}.

\begin{figure}[htb]
    \centering
    \includegraphics[width=0.4\linewidth]{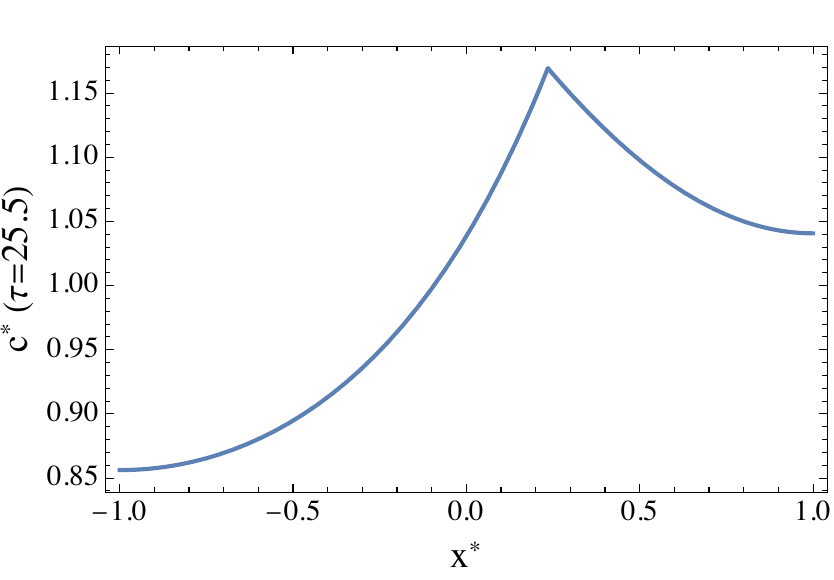}
    \caption{\small {A snapshot of the chemical concentration field released by the particle, $c^*(x^*,\tau)$ for $\tau=25.5, \mu^*=1,\lambda^*=5$.}}
    \label{fig:fieldsnapshot}
\end{figure}

 The dimensionless chemical field released by the particle can also be reconstructed numerically by summing over the field modes $$c^*(x^*,\tau) = \sum_{n=0}^N \alpha_n(\tau) \cos{\big(k_n(x^*+1)\big)}.$$ The chemical field evolves with the motion of the particle and a video of typical field dynamics for $\mu^*=1,\lambda^*=5$ is included in the supplementary material. A snapshot of this field distribution at time $\tau= 25.5$ is shown in Fig.~\ref{fig:fieldsnapshot}. The field exhibits a characteristic discontinuity in its first derivative at the particle location, where the concentration is maximal, and decays exponentially on either side.

  The numerical simulations reproduce the location of the phase transition with high accuracy. In Fig.~\ref{fig:phasediagram}\textcolor{blue}{(a)}, the green circular markers denote simulation runs with parameter values $\{\mu^*, \lambda^* \}$ that yield sustained oscillations. In contrast, the red crosses indicate simulations that exhibit damped oscillations that eventually relax to the stationary solution, $X^*_\tau = 0$. Comparison with the analytical line of stability $\{ \mu^*_c, \lambda^*_c\}$ demonstrates excellent agreement: the theory perfectly predicts the phase transition threshold for $\mu^* \in {[0,20]}$. Fig.~\ref{fig:phasediagram}\textcolor{blue}{(b)} shows that the theoretically predicted oscillation frequencies at the onset of the transition, $\omega^*_c$, are also well-captured by the numerical simulations in this range. For $\mu^* \gtrsim 20$, the numerical simulations become unstable. 
\begin{figure}
    \centering
    \captionsetup[subfigure]{labelformat=empty} 
    \begin{subfigure}{0.45 \textwidth}
        \centering
        \includegraphics[width=\linewidth]{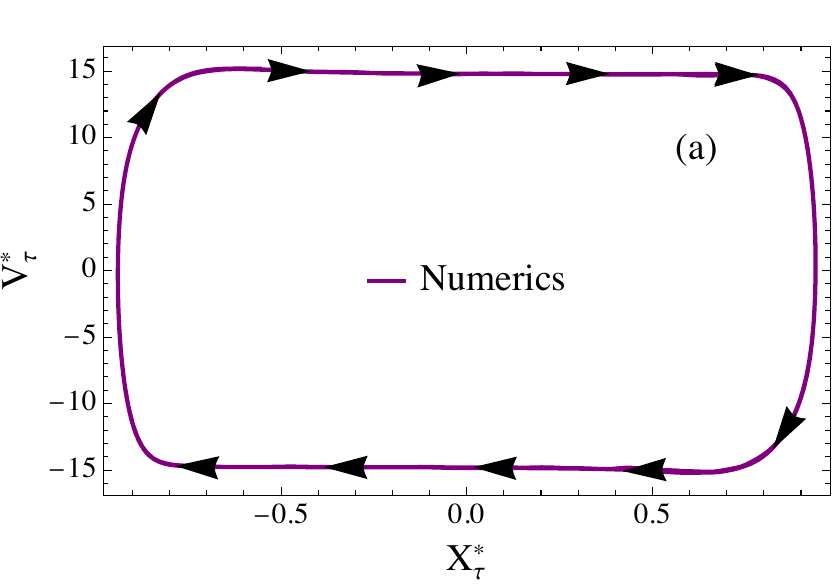}
    \end{subfigure}
    \hfill
    \begin{subfigure}{0.45 \textwidth}
        \centering
        \includegraphics[width=\linewidth]{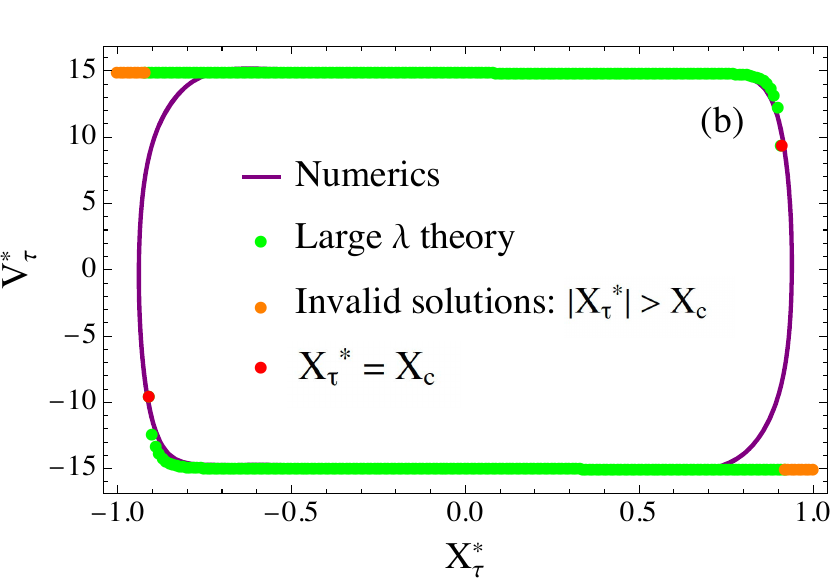}
    \end{subfigure}  
    \caption{\small {For $\mu^*=1, \lambda^* =30 $: \textbf{(a)} Numerical simulation phase portrait in the active regime far from the phase transition $(\lambda^*\gg \lambda_c^*)$. Black arrows indicate the particle's trajectory direction for increasing time. \textbf{(b)} Comparison of numerics with the theoretical prediction from the large $\lambda^*$ analysis. Green circular points represent valid numerical solutions to equation \eqref{j}}, orange points represent invalid solutions with $|X^*_\tau| >X_c$, and red points indicate the critical solution with $X_\tau^*=X_c$.}
    \label{fig:largeLa}
\end{figure}

 In Fig.~\ref{fig:largeLa}\textcolor{blue}(a) the non-elliptical limit cycle for parameter value $\lambda^* =30 \gg \lambda^*_c$ is shown to illustrate the steady-state particle behaviour far from the transition. The black arrows indicate the trajectory direction for increasing time. In this large $\lambda^*$ regime, the quasi-rectangular trajectories in the $(X_\tau^*,V_\tau^*)$ plane indicate that the particle exhibits translational motion with an almost constant velocity in the bulk of the channel, and rapidly decelerates to change direction when approaching the boundaries. For example, taking parameter values $\mu^*=1$ and $\lambda^*=30$, the mean velocity in the mid-channel region obtained by numerical simulation is $V^* = 14.8$. This is in excellent agreement with the analytical prediction for a free particle, $v^*=\sqrt{221}$, given by Eq.~\eqref{vroot} and setting D=1.
 
\begin{figure}[h]
    \centering
    \captionsetup[subfigure]{labelformat=empty} 
    \begin{subfigure}{0.45 \textwidth}
        \centering
        \includegraphics[width=\linewidth]{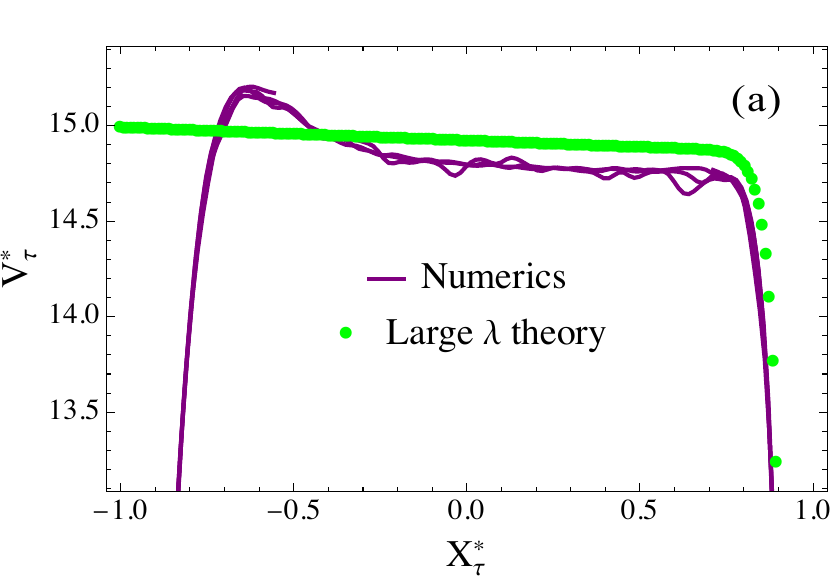}
    \end{subfigure}
    \hfill
    \begin{subfigure}{0.45 \textwidth}
        \centering
        \includegraphics[width=\linewidth]{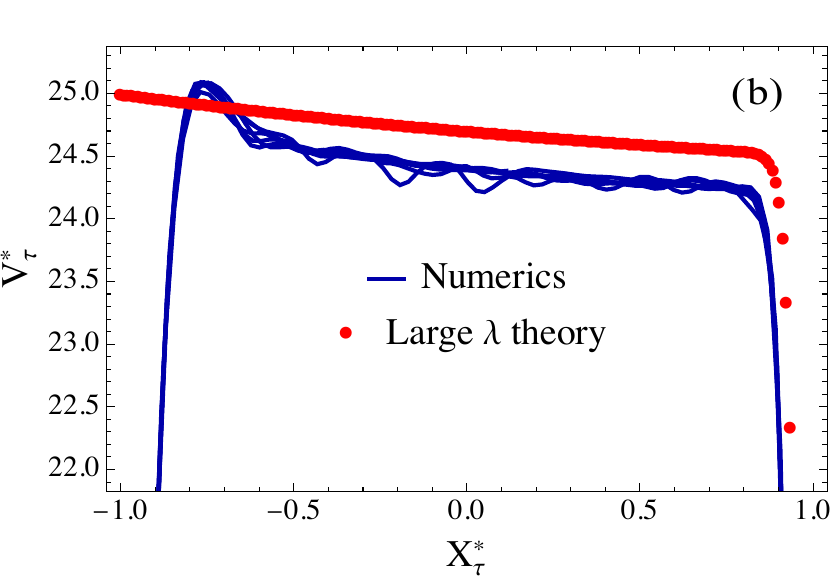}
    \end{subfigure}    

    \caption{\small { \textbf{(a)}: $\mu^*=1, \lambda^* =30 $: The purple line shows a zoomed-in portion of the numerics phase portrait in Fig.~\ref{fig:largeLa} and highlights the small increase in the particle velocity when reflecting near the boundary. The green circular points indicate the theoretical prediction for the large $\lambda^*$ regime. \textbf{(b)} $\mu^*=6, \lambda^*=50$: Numerics shown by the dark blue line and large $\lambda^*$ theory by red circular points. }}
    \label{fig:largeLtrajs}
\end{figure}

 In Fig.~\ref{fig:largeLa}\textcolor{blue}{(b)} we plot a phase trajectory comparison between numerics and the large $\lambda^*$ theory prediction, again for parameters $\mu^*=1,\lambda^*=30$. The green circular points indicate the theoretical prediction for the particle's coordinates $\{ X^*_\tau, V^*_\tau \}$ given by numerically solving equation \eqref{j} and setting $D=L=1$ to convert to dimensionless units. We highlight the critical point at $X^*_\tau= \pm X_c$ in red, beyond which real solutions to equation \eqref{j} vanish, as discussed in Sec.~\ref{sec:largelambdalimit}. Invalid numerical solutions with $|X_\tau^*| > X_c$ are highlighted in orange, and clearly do not represent valid coordinates in the physical trajectory due to the system's centrosymmetry. We find that the analytically predicted values from the large $\lambda^*$ theory agree remarkably well with the simulation data, even capturing the particle's deceleration  as it approaches the boundaries at $x^*= \pm 1$. The numerical simulation trajectory also indicates that the particle speed approaching the boundary is slightly lower than the speed leaving the boundary after reflection. Fig.~\ref{fig:largeLtrajs}\textcolor{blue}{(a)} displays a zoomed in portion of the phase space trajectory in Fig.~\ref{fig:largeLa}\textcolor{blue}{(b)} to highlight this asymmetry. Fig.~\ref{fig:largeLtrajs}\textcolor{blue}{(b)} shows a similar comparison between numerics and theory for parameter values $\mu^*=6$ and $\lambda^*=50$. 

 \begin{figure}[hbt!]
    \centering
    \captionsetup[subfigure]{labelformat=empty} 
        \begin{subfigure}{0.3\textwidth}
        \centering
        \includegraphics[width=\linewidth]{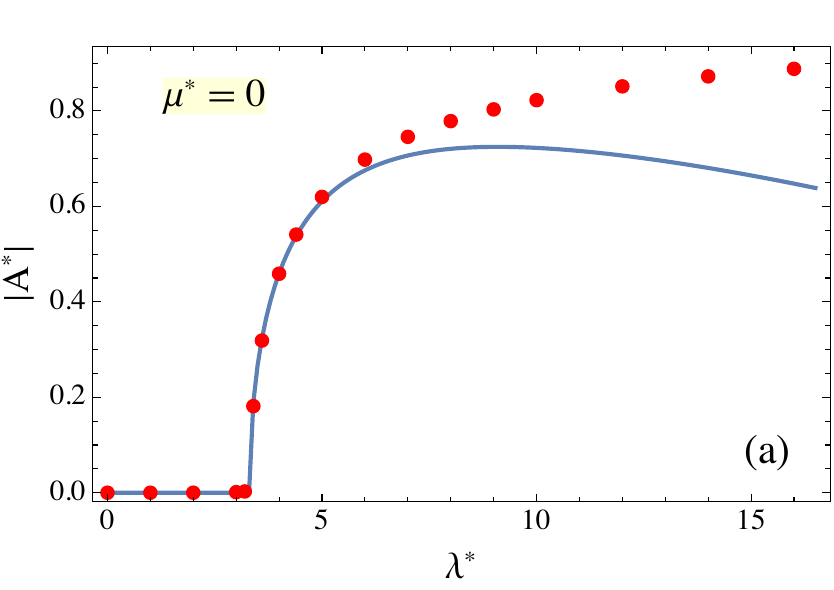}
    \end{subfigure}
    \hfill
    \begin{subfigure}{0.31\textwidth}
        \centering
        \includegraphics[width=\linewidth]{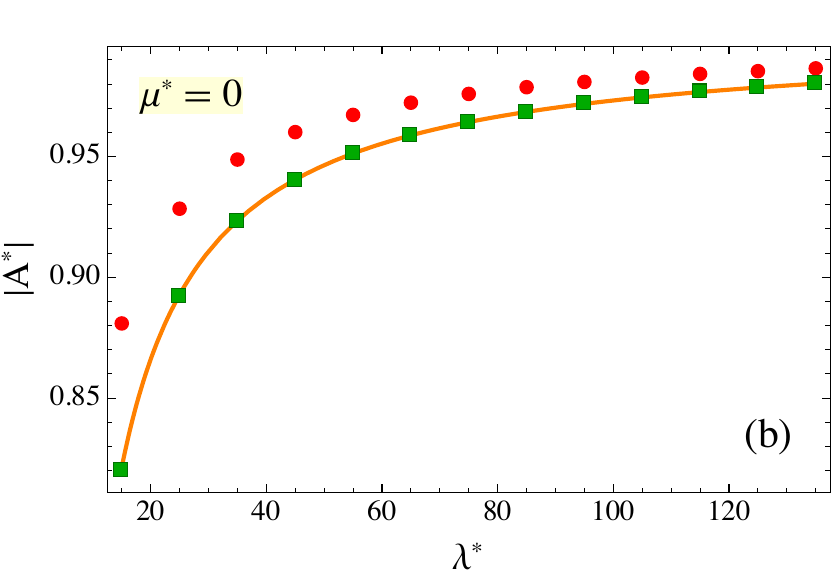}
    \end{subfigure}
   \hfill
    \begin{subfigure}{0.3 \textwidth}
        \centering
        \includegraphics[width=\linewidth]{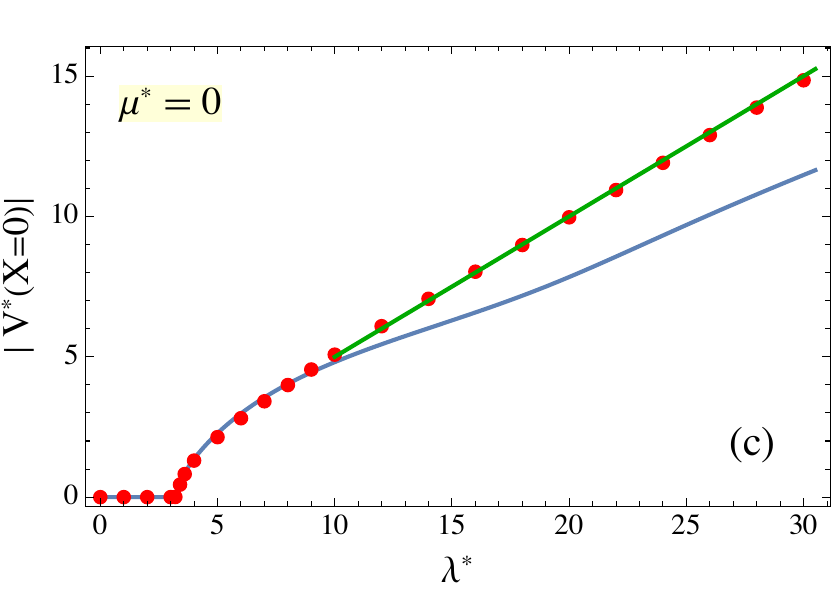}
    \end{subfigure}
    \begin{subfigure}{0.3\textwidth}
        \centering
        \includegraphics[width=\linewidth]{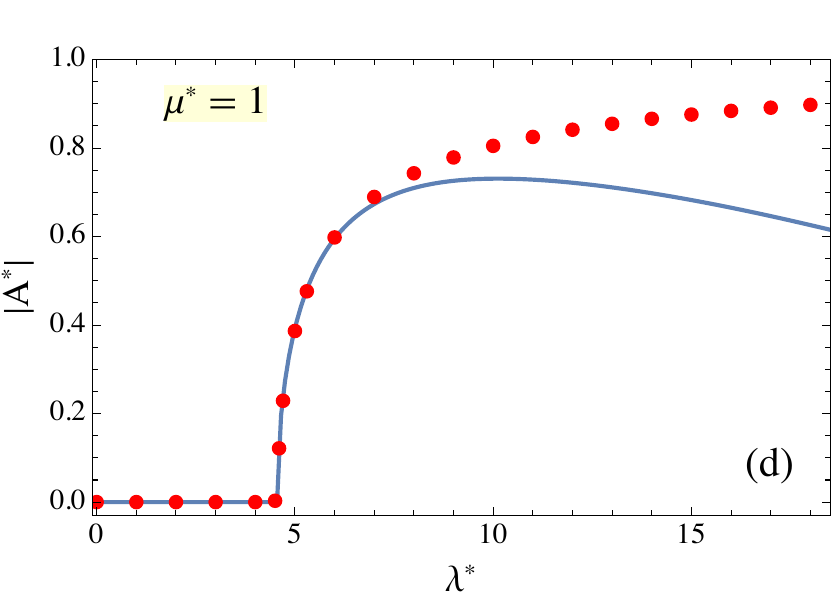}
    \end{subfigure}
    \hfill
    \begin{subfigure}{0.31\textwidth}
        \centering
        \includegraphics[width=\linewidth]{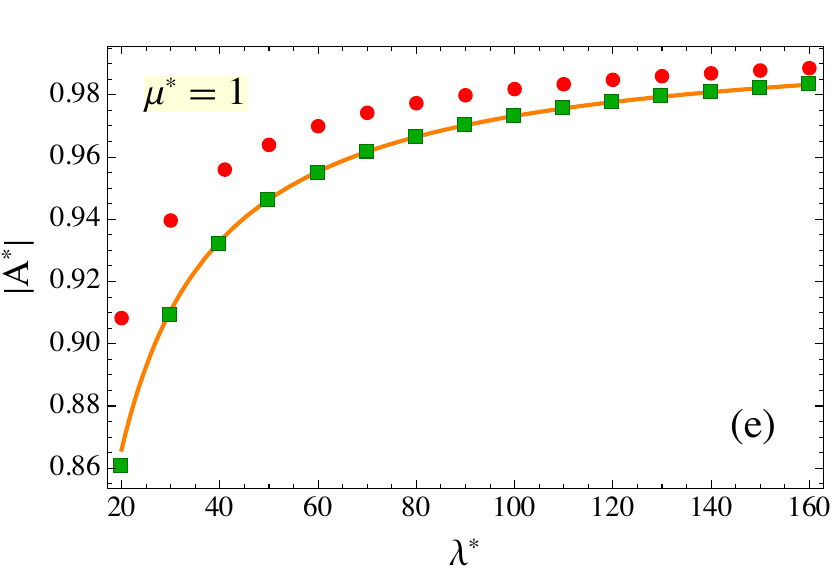}
    \end{subfigure}
       \hfill
    \begin{subfigure}{0.3\textwidth}
        \centering
        \includegraphics[width=\linewidth]{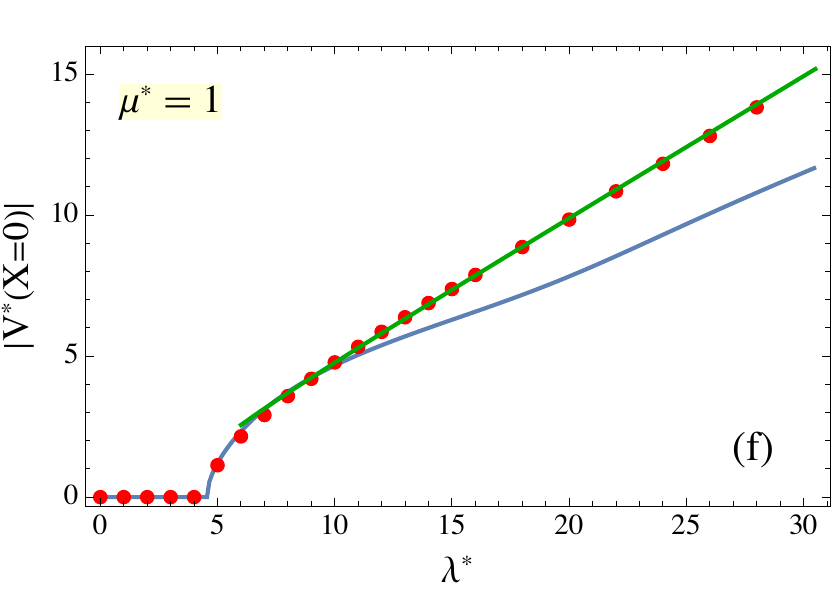}
    \end{subfigure}
    \begin{subfigure}{0.3 \textwidth}
        \centering
        \includegraphics[width=\linewidth]{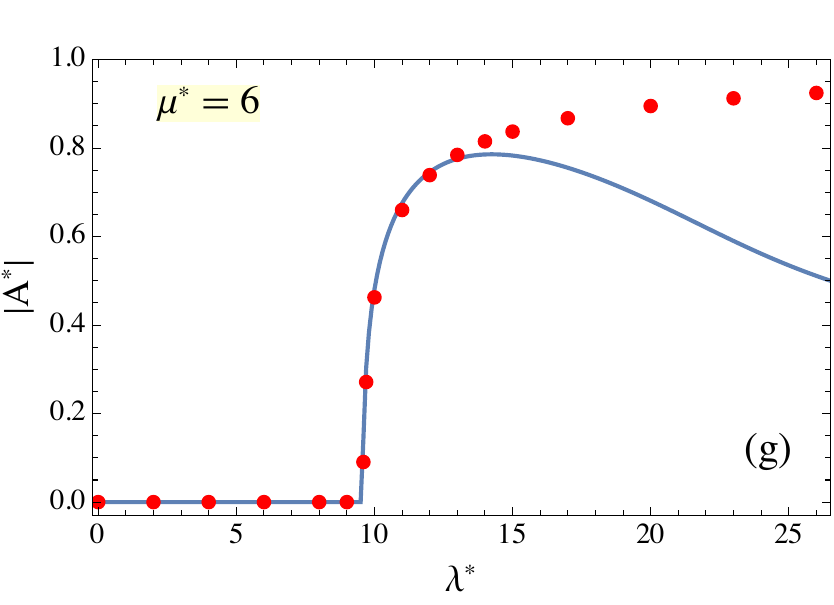}
    \end{subfigure}    
    \hfill
    \begin{subfigure}{0.31 \textwidth}
        \centering
        \includegraphics[width=\linewidth]{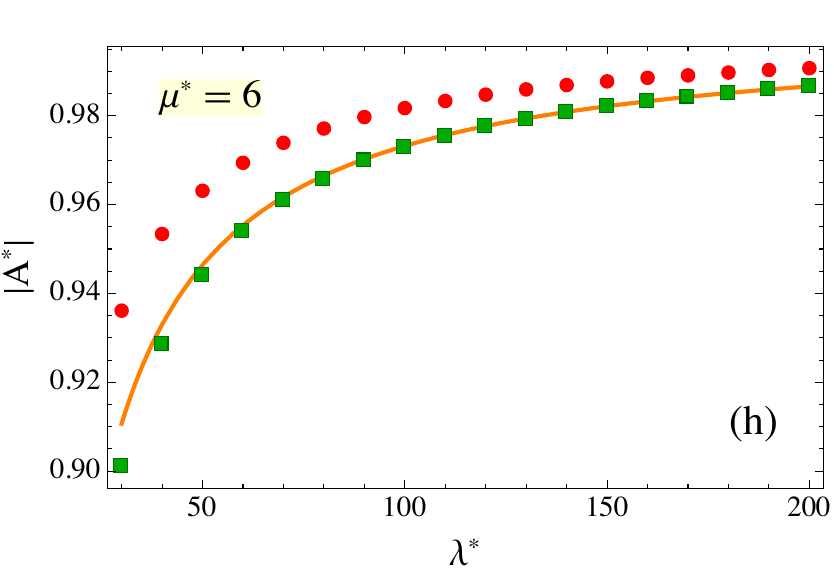}
    \end{subfigure}
       \hfill
    \begin{subfigure}{0.3 \textwidth}
        \centering
        \includegraphics[width=\linewidth]{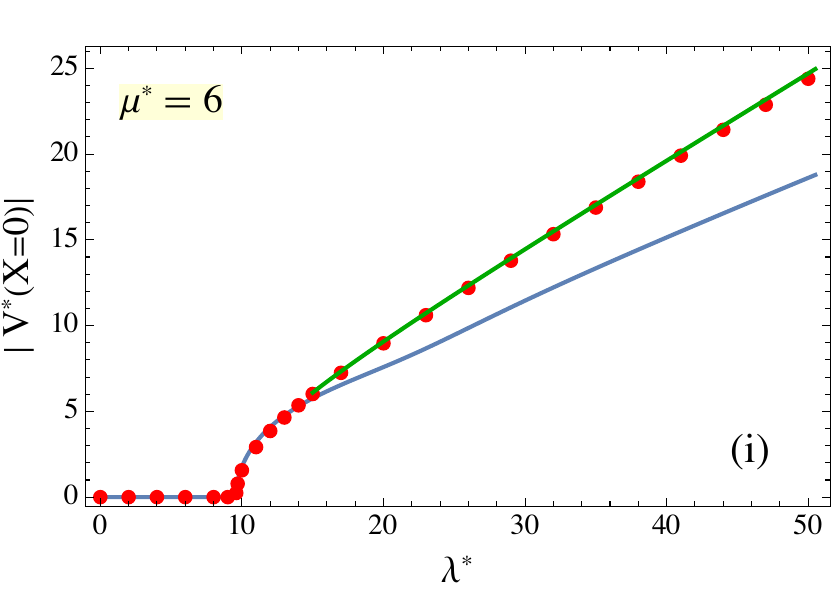}
    \end{subfigure}

    \vspace{0.5em}
    \includegraphics[width=0.8\textwidth]{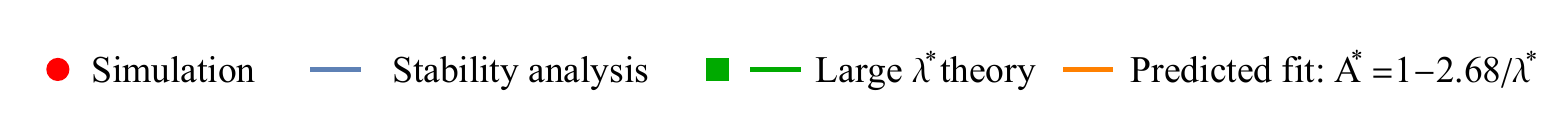}
    
     \caption{\small { Top row for $\mu^*=0$: $|A^*|$ as a function of $\lambda^*$ for $\lambda^*\sim\lambda^*_c$ in \textbf{(a)} and for $\lambda^* \gg\lambda^*_c$ in \textbf{(b)}. Stability analysis prediction is plotted as a blue line and simulation data with red circular markers. Predictions for $|A^*|$ from the large $\lambda^*$ theory are plotted with green square markers, while the orange line indicates the predicted behaviour $A^* = 1- \frac{2.68}{\lambda^*}$. In \textbf{(c)} $|V^*(X^*=0)|$ is plotted against $\lambda^*$. The plotted green line indicates the large $\lambda^*$ theory prediction. The corresponding plots for $\mu^*=1$ and $\mu^*=6$ are plotted in the middle (d-f) and bottom (g-i) rows, respectively.
     }}
     
    \label{fig:9boxes}
\end{figure}

 In Fig.~\ref{fig:9boxes} \textcolor{blue}{(a,d,g)} the particle's steady-state amplitude $|A^*|$ is plotted as a function of the phoretic mobility $\lambda^*$ in the regime close to the phase transition $(\lambda^* 
 \sim \lambda^*_c)$ and for respective evaporation rates $\mu^*=0,1,6$. The stability analysis prediction for $|A^*|$ is plotted as a blue line and the numerical simulation data with circular red points. For all values of $\mu^*$, the phase transition is clearly continuous with the amplitude of oscillations increasing smoothly from zero and following a pitchfork bifurcation characteristic of a supercritical transition. We see that the pitchfork bifurcation is extremely well captured numerically for $\lambda^* \sim \lambda^*_c$, and that the perturbation expansion holds remarkably well for oscillation amplitudes as large as $|A^*| \sim 0.7$. However, in the large lambda regime $(\lambda^* \gg \lambda^*_c)$ the perturbation expansion in $|X^*_\tau| \ll 1$ is no longer valid and the stability analysis solution for $|A^*|$ becomes unphysical. In this regime, a theoretical prediction for $|A^*|$ can be obtained by numerically solving $$V_\tau^* - \lambda^* \ \theta(X_\tau^*, V_\tau^*, \mu^*)  = 0,$$  where $\theta(X_\tau^*, V_\tau^*, \mu^*)$ is defined in \eqref{largeeV} with $D=L=1$. The predicted value of $|A^*|$ is then given by the maximum value of $X_\tau^*$ for which this equation admits real solutions for $V(X)$ with $V(X)>0$; beyond this point, such roots disappear. In Fig.~\ref{fig:9boxes} \textcolor{blue}{(b,e,h)} we plot this large $\lambda^*$ theory prediction for $(\lambda^*, |A^*|)$ with green square markers at respective evaporation rates of $\mu^* =0,1,6$. As predicted in Sec.~\ref{sec:largelambdalimit}, these points lie exactly on the curve $|A^*| =1 - \frac{2.68}{\lambda{^*}}$, which is plotted with an orange line. The red numerical simulation data converges on this large $\lambda^*$ prediction for increasing $\lambda^*$. In Fig.~\ref{fig:9boxes} \textcolor{blue}{(c,f,i)} the mean steady-state velocity at the mid-channel position, $|V^*(X^*_\tau=0)|$, is plotted as a function of $\lambda^*$ for $\mu^*=0,1,6$, respectively. The stability analysis prediction for the maximum velocity of the perfect harmonic oscillator, $|A^*\omega^*|$, is plotted as a blue line and is analogous to the particle velocity at the middle of the channel and in the regime close to the transition. Again, the red circular markers indicate numerical simulation values and show strong agreement with the stability analysis prediction for  $\lambda^* \sim \lambda^*_c$. The mid-channel velocity in the large $\lambda^*$ regime, as predicted by equation \eqref{largelambdV}, is also plotted with a green line and agrees well with the simulation data for $\lambda^* \gtrsim \lambda^*_c$.

\section{Discussion}\label{sec:discussion}

The results of Sec.~\ref{sec:results} demonstrate that the mathematical model studied here establishes a coherent theoretical framework for understanding confinement-induced active oscillations across a broad range of the dimensionless control parameters, $\{ \mu^*, \lambda^*\}$, describing the dynamics both in the weakly active regime and in the strongly active regime. In particular, by systematically analyzing the linear stability of the stationary state we obtain the  complete phase diagram separating passive and active regimes. The resulting stability boundary shows excellent agreement with numerical simulations throughout the explored parameter space. 

The results of our numerical simulations indicate that once the system enters the active phase, its dynamics converge toward a steady state characterized by self-sustained oscillations of the particle about the midpoint of the channel. In dynamical-systems terms, the long-time behavior is governed by a stable limit cycle. Remarkably, for a system with an infinite number of dynamical parameters (particle position plus field modes) there is no indication of any chaotic behavior. On this attractor, energy injection from the active phoretic mechanism is exactly balanced by dissipative losses, resulting in persistent oscillatory motion despite the absence of any underlying Hamiltonian structure. This behavior is reminiscent of the Rayleigh \cite{rayleigh} and Van der Pol \cite{VanderPol} oscillators, which are both paradigmatic examples of non-Hamiltonian systems in classical nonlinear dynamics. The former model system was first introduced by Lord Rayleigh to model self-excited acoustic oscillations, while the Van der Pol oscillator first  was proposed to model nonlinear, dissipative electrical circuits. Such systems broadly highlight the emergence of organized motion in driven, out-of-equilibrium systems. Indeed, the non-linear analysis of Koyano et al. \cite{Koyano1} for a camphor grain confined in a water channel is based on a perturbative expansion in position and its derivatives, which gives equations reminiscent of the Rayleigh oscillator \cite{rayleigh}.

A notable outcome of this work is the unexpected accuracy of the perturbative analysis based on small oscillation amplitudes. Although the expansion formally assumes $X_\tau^* \ll 1$, we find that the resulting predictions for the limit-cycle dynamics remain quantitatively reliable for amplitudes as large as $X_t \simeq 0.8$ (see Fig.~\ref{fig:9boxes}\textcolor{blue}{(g)}). This extended regime of validity is evidenced by the close agreement between analytical predictions and numerical trajectories, even well beyond the near-threshold region. The robustness of this perturbative approach suggests that the dominant physical mechanisms governing the oscillatory dynamics are already captured at low order, and that higher-order corrections play a comparatively minor role over a wide portion of parameter space.

Numerical simulations reveal that in the vicinity of the phase transition $(\lambda^* \sim \lambda^*_c)$ the limit cycle trajectory $\{ X_\tau^*, V_\tau^* \}$ is approximately elliptical, reflecting the near-harmonic nature of the oscillations in this regime. However, as the control parameter is tuned further into the active phase $(\lambda^*\gg \lambda^*_c)$, the shape of the limit cycle progressively deviates towards a more rectangular trajectory (with rounded edges), with an almost constant translational velocity in the bulk of the channel and rapid deceleration when reflecting at the boundaries. As previously reported by Koyano et al. \cite{Koyano1}, we find that the particle velocity in the region far from the boundaries is almost the same as that of a free particle in an infinite system. Therefore, the motion in the large $\lambda^*$ regime is regarded as the combination of translation in an infinite system and reflection by the boundaries. 

In the regime far from the phase transition we also observe a slight asymmetry in the trajectories: the particle rebounds from the boundary with a systematically higher speed than its speed of approach. As the particle approaches a boundary, the field it emits ahead of its motion is constrained by the no-flux boundary condition, leading to a localized accumulation of field intensity near the boundary. This accumulation generates a strong, short-ranged repulsive interaction that abruptly reverses the particle’s motion and causes the particle to depart from the boundary with a higher velocity than it had upon approach. The resulting speed asymmetry directly reflects intrinsic time irreversibility at the level of a single active particle. Such behavior cannot arise in equilibrium or Hamiltonian dynamics, where energy conservation and time-reversal symmetry forbid both stable limit cycles and asymmetric approach–departure velocities.

Finally, by examining the equations of motion in the limit of large phoretic coupling $\lambda$, we derive analytical results that characterize the particle dynamics deep in the active phase. In this strongly driven regime, the motion is dominated by intense, short-ranged interactions with the boundaries, leading to highly non-sinusoidal oscillations and pronounced velocity asymmetries. Nevertheless, the analytical expressions obtained in this limit accurately predict key features of the particle trajectories observed in simulations. Taken together with the near-threshold analysis, these results imply that the particle’s behavior can be understood and quantitatively predicted over nearly the entire $\{\mu^*,\lambda^* \}$ parameter space.

\section{Conclusion}\label{sec:conclusion}
We have analysed a minimal model for a self-phoretic particle confined in a one-dimensional channel and shown that geometric confinement alone can induce a transition from a passive stationary state to an active oscillatory regime. The transition occurs via a Hopf bifurcation when the particle–field coupling exceeds a critical value, leading to stable, self-sustained oscillations about the channel midpoint. The mechanism underlying this instability is the nonlinear feedback between the particle and its self-generated chemical field, whose confinement enhances memory effects and destabilizes the static solution.

By reducing the coupled particle–field dynamics to an effective equation of motion, we obtained exact analytical expressions for the phase boundary, the critical oscillation frequency, and the amplitude of oscillations near the transition. These predictions are in excellent quantitative agreement with numerical simulations over a wide range of parameters. Far from the bifurcation, where the oscillation amplitude becomes comparable to the system size, we derived complementary analytical results in the strong-coupling limit that accurately capture the particle’s boundary-dominated dynamics and velocity asymmetries. Taken together, these results provide a unified description of the system’s behavior across nearly the entire parameter space.

The present work highlights how confinement fundamentally alters self-phoretic dynamics by transforming a translational instability into a robust oscillatory state. Beyond its relevance to camphor-driven and other chemically active particles, the theoretical framework introduced here is readily extensible to other geometries, including periodic domains and higher-dimensional confined systems. An important direction for future work is the extension to multiple interacting particles, where coupling through shared chemical fields may give rise to collective oscillatory modes and synchronization phenomena. More generally, our results provide a controlled setting for studying how self-interactions and boundaries shape the dynamics of active matter at the single-particle level.

\section{Acknowledgments}
L.A. acknowledges support from the LIGHT S\&T Graduate Program (PIA 3 Investment for Future Program, ANR-17-EURE-0027) and D.S.D acknowledges support from the Grant No. ANR-23-CE30-0020-01 EDIPS . We also thank Thomas Gu\'erin, Yoann de Figueiredo and Antoine Aubret for enlightening discussions.

\section{Appendix: Image Charge Calculation}\label{sec:image}
In this section, we outline how the large $\lambda$ results of Sec.~(\ref{sec:largelambdalimit}) can be derived in terms of an image particle calculation. We consider here an unconfined phoretic particle on an infinite, one-dimensional domain. The reflecting boundary conditions at $x=\pm L$ can then be superficially imposed by introducing  image particles outside the physical domain $[-L,L]$. The derivative at $x=L$ can be made to vanish by placing an image particle at $x=2L-X_t$ in the interval $[L,3L]$. The boundary condition at $x=-L$ can then be imposed by reflecting this two-particle system on $[-L,3L]$ about $-L$ and $3L$, and so on. This generates image particles at the points $x=2mL +(-1)^{m} X_t$, where $m$ is an integer. The infinite series of source terms is thus given by $\sum_{m}\delta \big(x- 2mL -(-1)^{m}X_t \big)$, which can be decomposed in terms of m even and odd to give
 \begin{equation}
\sum_{n=-\infty}^\infty \delta(x- X_t+ 4nL ) +   \delta(x+X_t +2 L + 4 nL).
\end{equation}
In this model, the diffusion equation for the chemical field $c(x,t)$ can then be written on $\mathbb{R}$ as
\begin{equation}
\frac{\partial c(x,t)}{\partial t} = c''(x,t)  -\mu c(x,t)+\sum_{n=-\infty}^\infty \delta(x- X_t + 4nL  ) +\delta(x+ X_t + 2L+ 4n L),
\end{equation}
where we set $D=1$  for notational simplicity. We can now write a solution coming from each of the image sources moving in the same direction as the particle
\begin{equation}
\frac{\partial c_{+n}(x,t)}{\partial t} = c_{+n}''(x,t) -\mu c_{+n}(x,t)+ \delta(x- X_t+ 4nL  ),\label{c+eq}
\end{equation}
and those moving in the opposite direction to the particle 
\begin{equation}
\frac{\partial c_{-n}(x,t)}{\partial t} = c_{-n}''(x,t) -\mu c_{-n}(x,t)+ \delta(x+ X_t+ 2L + 4nL  ).
\label{c-eq}
\end{equation}
The total field is therefore given by the linear combination $c(x,t) = \sum_{n=-\infty}^\infty c_{+n}(x,t)+ c_{-n}(x,t)$.
Now we make the change of variable $y:= x- X_t + 4nL$ in Eq.~\eqref{c+eq} to find
\begin{equation}
\frac{\partial c_{+n}(y,t)}{\partial t}-\dot X_t \frac{\partial c_{+n}(y,t)}{\partial y} = c_{+n}''(y,t)-\mu c_{+n}(y,t) + \delta(y  ).\label{c+}
\end{equation}
Now, assuming that $\dot X_t=v$ can be treated as constant and that the time derivative can be neglected, we find $c_{+n}(y,t)= G_0(y,v)$ 
\begin{equation}
 v G'_0(y,v)+ G''_0(y,v) - \mu G_0(y,v)+\delta(y)=0.
\end{equation}
This implies that 
\begin{equation}
c_{+n}(x,t)= G_0(x-X_t+ 4 nL, v).
\end{equation}
The same argument can be used for equation \eqref{c-eq}, and one finds
\begin{equation}
c_{-n}(x,t)= G_0(x+X_t+ 2L+ 4 nL, -v).
\end{equation}
We therefore arrive at the approximate solution
\begin{equation}
c(x,t) \simeq   \sum_{n=-\infty}^\infty G_0(x-X_t +4nL,v) + G_0(x+X_t+4nL+2L,-v).
\end{equation}
The Green's function $G_0(y,v)$ has exponential solutions of the form
\begin{equation}
G(y,v)= A_\pm \exp \big(-\lambda_\pm(v) y \big),
\end{equation}
were $\lambda_+(v)$ is positive and $\lambda_-(v)$ is negative to assure convergence. The continuity and jump conditions at $y=0$ then give
\begin{equation}
A_+(v)=A_-(v),
\end{equation}
and 
\begin{equation}
-A_+(v)\lambda_+(v) +A_-(v)\lambda_-(v) +1 =0.
\end{equation}
So,
\begin{equation}
A_+(v)= \frac{1}{\lambda_+(v)-\lambda_-(v)}.
\end{equation}
This then gives for $y>0$
\begin{equation}
G_0(y,v) = \frac{\exp(-\lambda_+(v) y)}{\lambda_+(v)-\lambda_-(v)},\end{equation}
and for $y<0$
\begin{equation}
G_0(y,v) = \frac{\exp(-\lambda_-(v) y)}{\lambda_+(v)-\lambda_-(v)}.
\end{equation}
We have
\begin{equation}
-v\lambda +\lambda^2 -\mu =0,
\end{equation}
so
\begin{equation}
\lambda_{\pm}(v)=\frac{v \pm \sqrt{v^2 + 4\mu}}{2} \ {\rm for}\ v>0,
\end{equation}
and 
\begin{equation}
\lambda_{\pm}(-v)=\frac{-v \pm \sqrt{v^2 + 4\mu}}{2} \ {\rm for}\ v>0.
\end{equation}
Note from this we see that
\begin{equation}
\lambda_-(-v) =-\lambda_+(v); \ \lambda_+(-v) = -\lambda_-(v).
\end{equation}
Using these results we find the solution 
\begin{equation}
\begin{aligned}
&&c(x,X_t, v) \simeq  \frac{\exp \big(-\lambda_+(v) (x-X_t)\big)}{\big(\lambda_+(v)-\lambda_-(v)\big)\big(1- \exp(-4 L\lambda_+(v))\big)}+ \frac{\exp \big(-\lambda_-(v) (x-X_t)\big)}{\lambda_+(v)-\lambda_-(v)}\frac{\exp\big(4L\lambda_-(v)\big)}{1- \exp\big(4L\lambda_-(v)\big)}\nonumber \\
&+&\frac{\exp\big(\lambda_-(v) (x+X_t+2L)\big)}{\big(\lambda_+(v)-\lambda_-(v)\big)\big(1- \exp(4L\lambda_-(v))\big)}+ \frac{\exp\big(\lambda_+(v) (x+X_t-2L)\big)}{\big(\lambda_+(v)-\lambda_-(v)\big)\big(1- \exp(-4L\lambda_+(v))\big)}.
\end{aligned}
\end{equation}
The equation for $v$ as a function of $X_t$ is then given by
\begin{equation}
v(X_t) =-\lambda  \frac{\partial}{\partial x}c(x,X_t, v)|_{x=X_t}.
\end{equation}
Where one has to  symmetrize the discontinuous term in the derivative as discussed in the main text. This equation can then be written as 
\begin{equation}
v(X_t) = \lambda \theta(X_t,v(X_t)),
\end{equation}
and we find that,
\begin{equation}
    \begin{aligned}
\theta(X_t,v(X_t)) =-\frac{1}{2S} \bigg[&\frac{v\big(e^{2S} - e^{-2S} \big) + S\big( e^{-2v} - e^{2v}\big) +(v+S)e^{X_t(v+S)}\big[ e^{v-S} - e^{-(v-S)}\big]}{e^{2v} - e^{2S} - e^{-2S} +e^{-2v}}  \\ & \frac{ - (v-S)e^{X_t(v-S)} \big[e^{S+v} - e^{-(S+v)}\big] }{e^{2v} - e^{2S} - e^{-2S} +e^{-2v}} \bigg],
    \end{aligned}
\end{equation}
 with $S(v) \equiv \sqrt{v^2 + 4\mu}.$ This can be written as
 \begin{equation}
     \theta(X_t,v(X_t)) =\frac{1}{4S}  \bigg [ \frac{(v-S) e^{X_t(v-S)}}{ \sinh{(v-S)}}-\frac{(v+S) e^{X_t(v+S)}}{ \sinh{(v+S)}}  + (v+S) \coth{(v+S)} + (v-S) \coth{(S-v)}\bigg],
 \end{equation}
which is \eqref{largeeV} in the main text (setting $D=L=1$). 
\bibliographystyle{apsrev4-2}
\bibliography{references.bib}

\end{document}